\documentclass[iop,numberedappendix]{emulateapj}

\usepackage{color}

\begin{document}

\font\tenbg=cmmib10 at 10pt
\def \rvecxi{{\hbox{\tenbg\char'030}}}
\def \rvecphi{{\hbox{\tenbg\char'036}}}
\def \rvecdelta {{\hbox {\tenbg\char'016}}}
\def \rvecepsilon {{\hbox {\tenbg\char'017}}}
\def \rvecmu{{\hbox{\tenbg\char'026}}}
\def \rvecOmega {{\hbox {\tenbg\char'012}}}

\title{The role of fast  magnetic reconnection on the radio and gamma-ray emission from the nuclear regions of  microquasars and low luminosity AGNs}

\author{L.H.S. Kadowaki\altaffilmark{1},
        E.M. de Gouveia Dal Pino\altaffilmark{1},
        and Chandra B. Singh\altaffilmark{1}}

\altaffiltext{1}{Department of Astronomy (IAG-USP),
                 University of Sao Paulo, Brazil;
                 luis.kadowaki@iag.usp.br; dalpino@iag.usp.br}

\begin{abstract}
Fast magnetic reconnection events can be a very powerful mechanism operating in the core region of microquasars and AGNs. In earlier work, it has been suggested that the power released by fast reconnection events between the magnetic field lines lifting from the inner accretion disk region and the lines anchored into the central black hole could accelerate relativistic particles
and produce the observed radio emission from microquasars and low luminosity AGNs (LLAGNs). Moreover, it has been proposed that the observed correlation between the radio emission and the mass of these sources, spanning $10^{10}$ orders of magnitude in mass, might be related to this process. In the present work, we  revisit this model comparing
two different fast magnetic reconnection mechanisms, namely, fast reconnection driven by anomalous resistivity (AR) and by turbulence \citep[as described in][]{lazarian_vishiniac_99}. We apply the  scenario above to a much larger sample of sources (including also blazars, and gamma-ray bursts - GRBs), and  find that LLAGNs and microquasars do confirm the trend above. Furthermore, when driven by turbulence, not only their radio but also their gamma-ray emission can be due to magnetic power released by fast reconnection, which may accelerate particles to relativistic velocities in the core region of these sources. Thus  the turbulent-driven fast reconnection model  is able to reproduce better the observed emission than the AR model. On the other hand, the emission from blazars and GRBs does not follow the same trend as that of the LLAGNs and microquasars, suggesting that the radio and gamma-ray emission in these cases is produced  further out along the jet, by another population of relativistic particles, as expected. 

\end{abstract}


\section{Introduction}
\label{sec:intro}

  Galactic black hole binary systems (also referred as microquasars) and active galactic nuclei (AGNs) often exhibit variability and quasi-periodic relativistic outflow ejections of matter that may offer important clues about the physical processes that occur in their inner regions, in the surroundings of the central black hole (BH). Several authors have been exploring for decades these phenomena both observationally and also by means of theoretical and numerical modelling \citep[see, e.g.,][for recent reviews]{dgdp_lazarian_05, remillard_mcclintock_06, mcKinney_blandford_09, fender_belloni_12, zhang_etal_14}. 
 
 A potential model to explain the origin of these ejections and the often associated radio flare emissions was proposed by \citet[][hereafter GL05]{dgdp_lazarian_05} for microquasars and extended to AGNs and young stellar objects by \citet[][herafter GPK10]{dgdp_etal_10} \citep[see also][]{dgdp_etal_10b}.
Their model invokes the interactions between the magnetosphere anchored into the central BH horizon \citep{blandford_znajek_77} and the magnetic field lines arising from the accretion disk.

\begin{figure}[t]
 \centering
 \includegraphics[width=0.5\textwidth] {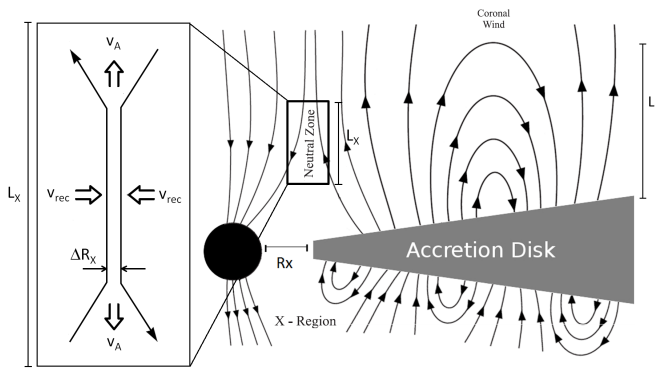}
 \caption{Schematic drawing of the magnetic field geometry in the region surrounding the BH. $R_{X}$ characterizes the inner accretion disk radius where the disk ram pressure balances the BH magnetosphere pressure. The lines arising from the accretion disk into the corona are pushed towards those anchored into the BH when the accretion rate is large enough and reconnection is fast at the magnetic discontinuity region of width $\Delta R_{X}$ and extension $L_X$ in the Figure (adapted from GL05). In the detail on the left side we highlight the magnetic reconnection site properties. 
In particular, when turbulence is present, this is superposed to the large scale magnetic field lines.}
 \label{fig:scheme}
\end{figure}

In accretion episodes where the accretion rate is increased (and may even approach the critical Eddington rate), both magnetic fluxes are pushed together in the inner disk region and reconnect under finite magnetic resistivity (see Figure \ref{fig:scheme}). In the presence of kinetic plasma instabilities \citep{shay_etal_04, yamada_etal_10}, anomalous resistivity \citep[e.g.,][]{parker_79, biskamp_etal_97, shay_etal_98}, or turbulence \citep[see][hereafter LV99]{kowal_etal_09, kowal_etal_12, lazarian_vishiniac_99}, reconnection becomes very efficient and fast (with reconnection velocities approaching the local Alfv\'en speed, which in these systems is near the light speed) and then may cause the release of large amounts of magnetic energy power. Part of this power will heat the coronal and the disk gas and part may accelerate particles to relativistic velocities. 

A first-order Fermi process for particle acceleration at the magnetic discontinuity was first described analytically in GL05 and then successfully tested numerically in collisionless pair plasmas by means of particle in cell simulations which can probe only the kinetic scales of this process \citep[see][]{drake_etal_06, drake_etal_10, zenitani_hoshino_08, zenitani_etal_09}\footnote{Where magnetic islands or Petcheck-like X-point configurations of fast reconnection are naturally driven and sustained by, e.g., the Hall effect or kinetic instabilities}, and more recently by means of 3D collisional MHD simulations with injected test particles which probed the efficiency of this process also at the macroscopic scales of the flow by \citet{kowal_etal_11, kowal_etal_12} \citep[see also][for reviews]{dgdp_etal_13, dgdp_etal_14}.
It has been found that particles accelerated by this process are able to produce Synchrotron radio spectra with power-law indices that are comparable to the observations \citep[][]{kowal_etal_12, dgdp_etal_13, dgdp_etal_14, delvalle_etal_14}\footnote{As argued in GL05, the particle acceleration mechanism  above is not the only possibility. Relativistic particles may be also produced behind shocks in the surrounds of the reconnection region. As in the Sun, plasmoids formed by reconnection of the field lines may violently erupt and cause the formation of a shock front and accelerated particles behind the shock front can also lead to power-law Synchrotron radio emission.}.  

Employing the magnetic reconnection model above in the surrounds of BHs, GPK10 found some evidence that the observed correlation between the radio luminosities and the BH source masses, spanning $10^{10}$ orders of magnitude in mass and $10^{6}$ in luminosity, from microquasars to low-luminosity AGNs (LLAGNs) \citep[see][]{merloni_etal_03, nagar_etal_05, fender_etal_04} could be explained by the magnetic power released by fast reconnection\footnote{By LLAGNs we mean sources with $L_{H\alpha} \leq 10^{40} erg~s^{-1}$ \citep[see][]{ho_etal_97, nagar_etal_05}.}. 
They also argued that this mechanism could be related to the transition between the observed ``hard''  and ``soft'' steep-power-law (SPL) states of microquasars \citep[e.g.,][]{remillard_mcclintock_06}.

Lately, similar mechanisms involving magnetic activity and reconnection in the core regions of compact sources to explain their emission spectra have been also invoked by other authors \citep[see, e.g.,][]{igumenshchev_09, soker_10, uzdensky_spitkovsky_14, dexter_etal_14, huang_etal_14}. In particular, magnetic reconnection between the magnetospheric lines of the central source and those anchored into the accretion disk resulting in the ejection of plasmons has been detected in numerical MHD studies \citep[see, e.g.,][]{romanova_etal_02, romanova_etal_11, zanni_ferreira_09, zanni_ferreira_13, cemeljic_etal_13}. The recent numerical relativistic MHD simulations of magnetically arrested accretion disks by \citet{tchekhovskoy_etal_11,mckinney_etal_12} and \citet{dexter_etal_14} also evidence the development of magnetic reconnection in the magnetosphere of the BH and are consistent with the scenario above. \citet{dexter_etal_14} also suggest that this process could be related to the transition from the ``hard'' to the ``soft'' SPL states and the transient phenomena observed in BH binaries. 

In the framework of very high energy (VHE) emission, for several years blazars (FSRQs and BL Lac objects), i.e., AGNs with highly beamed jets towards the line of sight, have been the largest sample detected by gamma-ray telescopes. This fact is consistent with the standard scenario for which the VHE of these sources is attributed to conventional relativistic particle acceleration along the jet, with the gamma-ray emission (due to either leptonic synchrotron self-Compton - SSC, or proton synchrotron, or Inverse Compton emission) being strongly Doppler boosted and producing the observed apparently high fluxes.

Recently, however, a few nearby radio galaxies, which are considered LLAGNs, have been also detected at TeV energies by HEGRA, HESS, MAGIC and VERITAS, namely M$87$, Cen A, and NGC$1275$ \citep[see, e.g.,][and references therein]{sol_etal_13}\footnote{There is also the case of IC$310$, a radio galaxy which has been interpreted as belonging to the blazar family, and a few other cases of radio galaxies to be confirmed at VHE}. The viewing angle of the large scale jets of these sources is of several degrees \citep[e.g., for M$87$ it is of the order of $30^o$;][]{reynoso_etal_11}, which allows only moderate Doppler boosting and the AGN source is highly underluminous (e.g., for M$87$, $L_{bol} \leq 10 - 5 L_{Edd}$). Both characteristics make it very hard to explain the VHE of these sources within the same standard scenario of blazars. Besides, the TeV emission in these sources is highly variable with timescales of a few days ($t_{var} \sim 1 - 2$ days for M$87$), pointing to extremely compact emission regions \citep[corresponding to scales of only a few Schwarzschild radii; e.g.,][]{abramowski_etal_12}. These findings have motivated several new studies on alternative particle acceleration mechanisms \citep[see, e.g.,][for reviews]{dgdp_etal_14} and emission theories \citep[e.g.,][]{neronov_aharonian_07, rieger_aharonian_08, tavecchio_ghisellini_08a, tavecchio_ghisellini_08b, abdo_etal_09b}, where gamma-ray GeV-TeV emission is considered to be produced in the vicinity of the BH (in a pulsar-like cascade mechanism) and/or at the jet launching basis. 

As described above, magnetic activity and reconnection events occurring close to BHs could also offer appropriate conditions for producing particle acceleration and the associated VHE gamma-ray emission in these sources, via interactions of the accelerated particles with the photon, density and magnetic fields in the surrounds of the BH. 
Similarly, microquasars are also expected to emit high energy gamma-rays, particularly owing to their general similarities to LLAGNs \citep[e.g.,][GPK10]{romero_etal_07}. Until now, only one source of this type, namely Cyg-X$3$, has been unambiguously detected in the GeV gamma-rays, by the Agile and Fermi observatories \citep{tavani_etal_09, abdo_etal_09a}. At TeV energies, only upper limit fluxes are available, in spite of intensive monitoring \citep[][]{aleksic_etal_10b}. There is also some evidence of sporadic GeV-TeV gamma-ray emission from Cyg X-$1$ \citep[][]{albert_etal_07, malyshev_etal_13} and upper limits in the $0.1-10$ GeV range for GRS $1915+105$ and GX $339-4$ \citep[see, e.g.,][]{bodaghee_etal_13}.
 
With regard to blazars, a recent study has identified an important correlation between these sources and gamma-ray bursts (GRBs) \citep[see][]{nemmen_etal_12}. These authors showed that the relativistic jets produced in both classes exhibit nearly the same correlation between the kinetic power carried out by accelerated particles and the gamma-ray luminosity. They concluded that this would be an indication that the bulk of the high energy emission comes from the jet region in these sources, therefore reinforcing the standard scenario for blazar emission. On the other hand, so far there has been no evidence of such a correlation for LLAGNs. This lack of correlation may be a further indication that the acceleration zones responsible for the observed high energy emission in the latter are not in the jet, but in the core, at the jet launching region, as suggested in the GL05 (and GPK10) model and by the observations described above. 

Since the gamma-ray emission is correlated with the accelerated relativistic particles that produce the radio observed emission (via, e.g., Inverse Compton, Synchrotron self-Compton, proton-proton, or proton-photon up-scatterings), in the present work we apply the GL05 scenario of reconnection-driven acceleration in the magnetized corona around the accretion disk and the BH of the sources and investigate whether or not the gamma-ray emission of microquasars and LLAGNs can be also associated to this mechanism in the core region. Moreover, we explore two possible  mechanisms for driving fast magnetic reconnection, namely, anomalous resistivity which was also employed in the previous GL05 and GPK10 studies, and turbulence based on LV99 model. We further extend the earlier study by GPK10, who found a correlation between the observed radio luminosity of the accelerated particles and the magnetic power released by reconnection in the core region for a few microquasars and LLAGNs, to a much larger sample of sources. We find that this trend is confirmed also in the much larger sample. We finally compare this with a sample of blazars and GRBs to examine the trend for higher luminous sources.

The sections of this work are divided as follows: in section \ref{sec:Wb} we briefly describe the model and summarize the main assumptions used to calculate the magnetic power released by fast magnetic reconnection in the core region of compact sources hosting BHs, comparing the results obtained for assuming either anomalous resistivity or turbulence to make reconnection fast. In section \ref{sec:comparison}, we describe the sample of sources and their observed core radio and gamma-ray emission which are compared with the theoretical predictions of our model. Finally, in section \ref{sec:conclusions}, we discuss the implications of our results for all classes of sources 
and draw our general conclusions.

\section{Magnetic power produced by fast reconnection in the surrounds of a BH}
\label{sec:Wb}

\subsection{Possible scenario for the accretion disk and corona right before fast magnetic reconnection}

Our model is described in detail in earlier work (GL05 and GPK10) and we summarize here its main assumptions introducing important new upgrades as described below. As in GL05 and GPK10, 
we assume that the inner region of the accretion disk/corona system  alternates between two states which are controlled by changes in the global magnetic field. As described in the previous section, we consider that right before a fast magnetic reconnection event the system is in a state that possibly characterizes the transition from the hard to the soft state, and adopt a magnetized standard (geometrically thin and optically thick) accretion disk \citep{shakura_sunyaev_73} with a corona around a BH, as shown in the cartoon of Figure \ref{fig:scheme}. 
Nevertheless, we must stress that the real structure of the accretion disk is not a crucial point for the purposes of our study because our focus is the inner coronal disk region where the interaction of the magnetic field lines with the BH magnetosphere takes place (see more in the Discussion section).

A magnetosphere around the central BH may be built from the continuous drag of magnetic field lines by the accretion disk \citep[e.g.,][]{macdonald_etal_86, wang_etal_02}. The coronal disk large-scale poloidal magnetic field can be in turn established either by advection of lines carried from the outer regions of the disk or by the action of a dynamo inside the accretion disk \citep[e.g.,][]{livio_etal_03, king_etal_04, uzdensky_goodman_08, krolik_piran_11, krolik_piran_12} possibly driven by the combination of magnetorotational instability \citep[see][]{balbus_hawley_98} and disk differential rotation. 

According to mean field dynamo theory (see GL05 and references therein, an inversion of the polarization of the large scale magnetic lines is expected to occur every half of the dynamo cycle; when this happens a new flux of disk lines should reach the inner region with an inverted polarity with respect to the magnetic flux already anchored into the BH, therefore, favouring magnetic reconnection between the two fluxes (GL05, GPK10). 
Whether the magnetic flux is produced by a dynamo in the disk itself or advected from the outer regions, or both, its sign is expected to change periodically, although the characteristic time scale of this variation is hard to compute in the absence of a detailed modelling of the dynamo in the disk or of the flux advected from the outer regions \citep[][GL05]{tagger_etal_04, tchekhovskoy_etal_14}. 
In the case of the microquasar GRS $1915+105$, for instance, it has been suggested \citep[see][]{tagger_etal_04} that the long term evolution of the field configuration could be of the order of one to a few years so that half of the time the field fluxes in the inner disk edge and in the BH would be antiparallel and the rest of the time parallel. \citet{tchekhovskoy_etal_14} suggest that the time scales  might be possibly regulated by the accretion time-scale at the outer radii of the disk ($t_{acc} = \alpha^{-1} (H/r)^{-2} \Omega_K^{-1}$, where $\Omega_K= (GM/ r^3 )^{1/2}$, with $\alpha$ being the disk viscosity parameter, $H$ its height and $r$ the radial distance) which may imply much faster time scales. Numerical simulations of MRI-driven MHD turbulence, which involves part of a mean field dynamo growth, show that generally $\sim 10$ orbital periods are required to reach saturation \citep[][]{stone_etal_96}, but the building of the large scale poloidal component may take longer and thus the time for its inversion in the disk. \citet{dexter_etal_14} performed 3D MHD long term simulations letting  material to be accreted with a magnetic field with opposite polarity with respect to that   attached to the BH and detect a field inversion in the later after $t \sim 2 \times 10^4 R_S/c$ (where $R_S$ is the Schwarzschield radius), which also indicates a very fast process. All these processes are usually connected with different variability phenomena that is detected in BH sources and span a large interval of time scales. The determination of a precise characteristic time scale at which the system may reach exactly the configuration as idealized in Figure \ref{fig:scheme} is out of the scope of the present work. Nevertheless, the recent observations of a dynamically important magnetic field near the Galactic Centre black hole \citep[see, e.g.,][]{zamaninasab_etal_14}, as well as relativistic numerical simulations of accretion disk-BH magnetosphere interactions as described above \citep[see][]{mckinney_etal_12, dexter_etal_14} indicate that this is a quite possible configuration in the surrounds of BHs. 

The poloidal magnetic field lines built in the corona summed to the disk differential rotation give rise to a wind that removes angular momentum from the system leading to an increase in the accretion rate and thus an increase in the ram pressure of the accreting material. This will further accumulate the magnetic lines in the inner disk coronal region pressing them against the lines anchored in the BH horizon and facilitating the production of a fast magnetic reconnection event (see GL05 and the neutral zone in Figure \ref{fig:scheme}).
As shown in GL05 and GPK10, a fast magnetic reconnection event may release  substantial magnetic power. In order to evaluate this amount of magnetic power we need first to characterize the coronal parameters in the inner disk region. 

We  consider here a strongly magnetized fluid in the surrounds of the BH for which the condition $R_{i} < L_{mfp} < L$ is fulfilled (where $L$ characterizes the large scale dimension of the system, $L_{mfp} \sim 1.8 \times 10^{4} n^{-1} T^2 ~cm$ is the ion mean free path for Coulomb collisions,  and $R_i \sim 2.1 \times 10^9 (E/B) ~cm$ is the ion Larmor radius; see more below). For such scales a weakly collisional or effectively collisional MHD description is more than appropriate \citep[e.g.,][]{kulsrud_83}.  

It should be noticed also that we  consider a nearly non-relativistic MHD approach for describing the coronal region around the BH \citep[see also, e.g.,][]{liu_etal_03}. The ion/electron temperatures are smaller than or equal to $\sim 10^9$ K, which makes the fluid approximately non-relativistic and reasonably well described by the standard equations below. 
Nevertheless, with regard to reconnection, the fact that the Alfv\'en speed $v_A$ may become comparable to the light speed for the conditions analysed here may imply that eventually fast reconnection becomes nearly relativistic. Based on studies performed mostly for collisionless reconnection \citep[see, e.g., the reviews of][and references therein]{uzdensky_11, lyutikov_lazarian_13,dgdp_etal_14}, it has been found that the behaviour of slow and fast reconnection in relativistic regimes is compatible with that of non-relativistic reconnection. There is also a recent work by \citet{cho_Lazarian_14} where it is demonstrated that relativistic collisional MHD turbulence behaves as in the non-relativistic case. This indicates that collisional turbulent fast reconnection  theory (as described, e.g., in the LV99 model, see below) can be directly applicable to the relativistic case \citep[see also][]{lyutikov_lazarian_13} and we will adopt here this approach. 

As remarked, we  employ below the standard optically thick, geometrically thin accretion disk model \citep{shakura_sunyaev_73}.
In GL05 and GPK10 works, the inner radius of the accretion disk ($R_X$) was taken at the last stable orbit around the BH ($3R_S$, where $R_S = 2GM/c^2 = 2.96 \times 10^5 M/M_{\odot}~cm$ is the Schwartzschild radius). Although physically possible, this condition may lead to a singularity in the Shakura-Sunyaev disk solutions and therefore, we presently adopt an inner radius $R_X = 6 R_S$, which does not affect much the earlier results numerically but avoids the singularity. For a BH with stellar mass $M = 14 M_{\odot}$ (which is suitable for microquasars) this gives $R_X = 6 R_S \simeq 2.48 \times 10^7~cm$. 
Besides, this condition ensures that $R_X$ and all characteristic large scales of the system will satisfy the condition for employment of the nearly collisional MHD fluid approximation (namely, $R_{i} << L_{mfp} \lesssim R_{X}$).

In order to determine the magnetic field intensity in the inner region immediately before an event of violent magnetic reconnection, we assume the equilibrium between the disk ram pressure and the magnetic pressure of the BH magnetosphere.
As in GPK10, we approach the radial accretion velocity by the free fall velocity and also assume that
the intensity of the field that was dragged by the disk and anchored into the BH horizon neighbourhood is of the order of the inner disk magnetic field intensity \citep[see][and GL05]{macdonald_etal_86}, which gives

\begin{equation}
 {\dot{M}\over 4\pi R^2}\left({2GM\over R}\right)^{1\over2} \sim {B_{d}^2\over8\pi}.
 \label{eq:ram_pressure}
\end{equation}

or\footnote{We note that in GPK10, eq.(\ref{eq:ram_pressure}) was parametrized in terms of $\beta$, namely the ratio between the total disk pressure (gas $+$ radiation pressure) and the magnetic pressure, rather than in terms of $\dot{m}$. In the case of a radiation pressure dominated disk \citep{shakura_sunyaev_73}, both parameters are related through the equation: $\beta \simeq 0.12 \alpha^{-1} r_{X} \dot{m}^{-1}$.}

% equation ok!
\begin{equation}
  B_{d} \simeq 9.96 \times10^{8} r_{X}^{-{5\over4}} \dot{m}^{1\over2} m^{-{1\over2}}~~G,
  \label{eq:bd}
\end{equation}
where $r_{X}=R_{X}/R_{S}$ is the inner radius of the accretion disk in $R_{S}$ units,
$\dot{m}=\dot{M}/\dot{M}_{Edd}$ is the mass accretion rate in $\dot{M}_{Edd}$ units (which corresponds to the Eddington mass accretion rate $\dot{M}_{Edd} =1.45\times10^{18} m~g/s$), and $m=M/M_{\odot}$ is the BH mass in solar mass units.
 
To quantify the parameters of the corona right above the inner disk region, as in GL05 and GPK10 we 
employ the model of \citet{liu_etal_02}:

% equation ok!
\begin{equation}
 T_c \simeq 1.74\times10^{6} \Gamma^{1\over4}B_{d}^{3\over4} L^{1\over8} U_{rad}^{-{1\over4}}~~K,
  \label{eq:tcr}
\end{equation}
% equation ok!
\begin{equation}
 n_c \simeq 9.64\times10^{17} \Gamma^ {1\over2} B_{d}^{3\over2} L^{-{3\over4}} U_{rad}^{-{1\over2}}~~cm^{-3},
  \label{eq:ncr}
\end{equation}
where $T_c$ and $n_c $ are the coronal temperature and density, respectively, $L$ is the size of a coronal magnetic flux tube (which will also characterize the height of the corona), $U_{rad}$ is the disk soft radiation energy density. In GL05 and GPK10, to evaluate this quantity we neglected for simplicity the effects of disk opacity. Here, we give $U_{rad}$ in terms of the effective temperature at the disk surface \citep[e.g.,][]{frank_etal_02, liu_etal_03}:
 
% equation ok!
\begin{equation}
U_{rad}=aT_{eff}^4={4\over c}{3GM\dot{M}q^4 \over 8 \pi R_{X}^3}.  
  \label{eq:urad}
\end{equation}    
where $q=[1-(3R_{S}/R_{X})^{1/2}]^{1/4}$.

 Using equations (\ref{eq:bd}) and (\ref{eq:urad}), the coronal parameters can be rewritten as

% equation ok! 
\begin{equation}
  T_{c}= 2.73\times10^{9} \Gamma^{1\over4} r_{X}^{-{3\over16}} l^{1\over8} q^{-1} \dot{m}^{1\over8}~~K,
  \label{eq:tcr02}
\end{equation}
\begin{equation}
  n_{c} \simeq 8.02 \times10^{18} \Gamma^{1\over2} r_{X}^{-{3\over8}} l^{-{3\over4}} q^{-2} \dot{m}^{1\over4} m^{-1}~~cm^{-3},
  \label{eq:ncr02}
\end{equation}  
where $l = L/R_{S}$. Also, instead of employing $v_A \simeq c$ as in GL05 and GPK10, we have replaced $v_A$ by its relativistic form $v_A = \Gamma v_{A0}$, with $v_{A0} = B/(4\pi \mu m_{H} n_{c})^{1/2}$,
$m_H= m_p$ is the proton rest mass, $\mu \sim 0.6$,   and  $\Gamma=[1+({v_{A0} \over c})^2]^{-1/2}$ \citep[e.g.,][]{somov_12}\footnote{We note that for the parametric space considered here, $0.36 \lesssim \Gamma \lesssim 0.99$ is obtained numerically from the solution of the equation $v_{A0}\simeq 9.78 \times 10^{10} \Gamma^{-{1\over4}}r_{X}^{-{17\over16}} l^{3\over8}q\dot{m}^{3\over8}$.}.

As described in GL05 and GPK10, the rate of magnetic energy that can be extracted from the reconnection (neutral) zone in the corona (above and below the disk in Figure \ref{fig:scheme}) through reconnection is $\dot{W}_{B} = (B^2/ 8\pi) \xi v_{A} (4 \pi R_{X} L_{X})$, where $\xi = v_{rec}/v_A$ is the magnetic reconnection rate, $v_{rec}$ is the reconnection velocity, and $L_{X}$  is the length of the reconnection region. Mass flux conservation and Figure \ref{fig:scheme} imply that $\Delta R_X/L_X = v_{rec}/v_{A}$. Therefore,

% equation ok!
\begin{equation}
 \dot{W}_{B} = {B^2 \over 8\pi} v_{A} (4 \pi R_{X} \Delta R_{X}),
  \label{eq:wb00}
\end{equation}
where $B$ is the coronal magnetic field in the reconnection zone which is of the order of $B_d$, and $\Delta R_{X}$ is the width of the current sheet. Its estimate depends on the fast reconnection model adopted. 

\subsection{Fast Reconnection driven by anomalous resistivity}

In GL05 and GPK10 we assumed that fast reconnection is driven by anomalous resistivity \citep[see][LV99]{parker_79, biskamp_etal_97}.  
This is based on the onset of current driven instabilities, that can enhance the microscopic Ohmic resistivity and speed up reconnection \citep[e.g.,][]{papadopoulos_77, parker_79, biskamp_etal_97}. It gives rates which are much faster than the standard Sweet-Parker slow reconnection (which is driven by Ohmic resistivity) and may be naturally present in regions of strong magnetic fields, like those around the BH.
When the electron-ion drift velocity exceeds the electron thermal velocity, there can be an electron runaway which causes the formation of electron beams that in turn generate plasma electrostatic waves, giving rise to collective interactions. Electrons are scattered by these fields rather than by individual ions and the classical Spitzer resistivity is replaced by an anomalous resistivity \citep[e.g.,][]{papadopoulos_77}. Following \citep[][LV99, see also GL05]{parker_79, biskamp_etal_97}, we can estimate the width of the current sheet for anomalous resistivity:
  
% equation ok!
\begin{equation}
\Delta R_{X}=\frac{c\Delta B}{4\pi n_{c} Z e v_{th,c}},
\label{eq:deltar00}
\end{equation}
where $\Delta B \simeq 2B \simeq 2B_{d}$, and $v_{th,c}=(kT_{c}/m_{p})^{1/2}$ is the thermal velocity of the ions of charge $Ze$ in the corona. From equations (\ref{eq:bd}), (\ref{eq:tcr02}) and (\ref{eq:ncr02}), we obtain:

% equation ok!
\begin{equation}
 \Delta R_{X} \simeq 2.02 \Gamma^{-{5\over8}}r_{X}^{-{25\over32}} l^{11\over16} q^{5\over2} \dot{m}^{3\over16} m^{1\over2}~~cm.
  \label{eq:deltar}
\end{equation}

 Under these conditions, the magnetic energy power released during violent fast magnetic reconnection in the surrounds of a BH is approximately given by\footnote{We note that we have also explored the effects of general relativity in the gravitational potential in the surrounds of the BH and recalculated the magnetic reconnection power in the inner disk region considering a pseudo-Newtonian potential and found that for the scales considered here with $R_X \simeq 6 R_S$ these are negligible.}

% equation ok! 
\begin{equation}
 \dot{W}_B \simeq 2.89 \times10^{34} \Gamma^{1\over8} r_{X}^{-{107\over32}} l^{17\over16} q^{7\over2} \dot{m}^{25\over16} m^{1\over2}~~erg/s.
  \label{eq:Wb}
\end{equation}

We note that with the adopted normalization, $\Gamma$, $r_{X}$, $l$, $q$, $\dot{m}$, and $m$ are dimensionless parameters.

\subsection{Fast magnetic reconnection driven by turbulence}
\label{sec:Wbturb}

In the previous section, we derived the magnetic power released by fast magnetic reconnection due to  anomalous resistivity \citep[see][]{parker_79, biskamp_etal_97}. Eq. (\ref{eq:deltar}) shows that the resulting thickness of the reconnection region is very small compared to the large scales of the system. Though still much larger than the ion Larmor radius ($R_i \simeq 6.4 \times10^{-5}\Gamma^{1/8} r_{X}^{37/32}l^{1/16} q^{-1/2} m^{1/2}\dot{m}^{-7/16}~cm$), it indicates that anomalous resistivity prevails only at the small scales of the system. 

On the other hand, the systems we are dealing with (microquasars and AGNs), as most astrophysical systems, have very large Reynolds number, $R_{e} \sim LV/\nu \sim 10^{20} - 10^{28}$, where $V$ is a characteristic fluid velocity, $\nu$ is the viscosity which for a magnetically dominated fluid is mainly normal to the magnetic field and is  given by $\nu_{\perp} \simeq  1.7 \times 10^{-2} n \ln\Lambda T^{-1/2} B^{-2}~cm^{2} s^{-1}$,  where $\ln \Lambda$  is the Coulomb logarithm, 
$\Lambda = 3/2 e^3 (k^{3} T^{3} / \pi n)^{-1/2} \min[1,(4.2\times10^5/T)^{1/2}]$ \citep[see][]{zhang_yan_11, spitzer_62}.
Similarly, the magnetic Reynolds number $R_m = LV/\eta \sim 10^{18} - 10^{24}$, where $\eta$ is the magnetic resistivity which in the regime of strong magnetic fields is given by $\eta_{\perp} \simeq  1.3 \times 10^{13} Z \ln\Lambda T^{-3/2} ~cm^{2} s^{−1}$ \citep[][]{spitzer_62}.   
These large Reynolds numbers imply that the fluid and the magnetic fields can be highly distorted and turbulent if there is turbulence driving. In other words, the growth of any instability as for instance, the current driven instabilities mentioned above (like, e.g., the Buneman instability which occurs when $T_e/T_i \sim 1$, where $T_e$ and $T_i$ are the electron and ion temperatures, respectively, \citet{papadopoulos_77} can naturally drive turbulence with characteristic velocities around the particles thermal speeds. Also, the occurrence of continuous magnetic reconnection with the ejection of plasmoids during the building of the corona itself in the surrounds of the BH \citep{liu_etal_02,liu_etal_03} will contribute to the onset of turbulence. Numerical simulations of coronal disk accretion also indicate the formation of turbulent flow in the surrounds of the BH that may be triggered, e.g., by magnetorotational instability \citep[see e.g,][]{tchekhovskoy_etal_11, mckinney_etal_12, dexter_etal_14}. 

Turbulence is known to speed up the reconnection.
In this case, we can  examine an alternative model of fast reconnection driven by turbulence. We here adopt the turbulent collisional fast reconnection model introduced by LV99.

According to LV99 theory, the presence of even weak turbulence in a collisional flow causes the  wandering of the magnetic field lines (facilitated by Richardson diffusion, see below) which allows for simultaneous events of reconnection to occur in small patches making reconnection naturally fast and independent of the Ohmic resistivity. In other words, the spontaneous stochasticity introduced by the weak turbulence in the mean field causes the diffusion of the magnetic field lines, which is a macroscopic process independent of microscopic resistivity. The reconnection over small patches of magnetic field determines the local reconnection rate. The global reconnection rate is substantially larger as many independent patches reconnect simultaneously. In other words, the LV99 model predicts that the small scale events happen at a slow (Sweet-Parker) rate, but the net effect of several simultaneous events makes reconnection fast.
This theory has been thoroughly discussed  in the literature in many papers and reviews \citep[see, e.g.,][]{eyink_etal_11, eyink_etal_13, lazarian12, lazarian15} and successfully tested by means of 3D MHD simulations \citep{kowal_etal_09, kowal_etal_12, xu13}.

In the LV99 theory, the reconnection velocity is given by:

\begin{equation}
v_{rec} \simeq v_{A} \min\left[{L_{inj}\over L_X}, {L_X\over L_{inj}}\right]^{1\over2}M_{A}^2,
\label{eq:vrec_turb}
\end{equation}
where $L_X$ is the extension of the reconnection zone as in Figure \ref{fig:scheme}, $M_{A}={v_{inj}/v_{A}}$ is the Alfv\'enic Mach number of the turbulence, $v_{inj}$ and $L_{inj}$ are the turbulent velocity and length, respectively, at the injection scale. 
The three-dimensional numerical MHD simulations by \citet{kowal_etal_09,kowal_etal_12} indicated a slightly distinct dependence in the power-law index in the equation above, i.e., $v_{rec} \propto L_{inj}^{1/4}$. 

The equation above was also derived using the well-known concept of Richardson diffusion \citep{eyink_etal_11}. From the theoretical perspective this new derivation avoids the more complex considerations of the cascade of reconnection events that were presented in LV99 to justify the model.  These authors have demonstrated that the LV99  model is connected to the ``spontaneous stochasticity'' of the magnetic field in turbulent fluids \citep[see also][where numerical simulations demonstrate this diffusion process as well]{eyink_etal_11}.

Mass conservation ($\Delta R_{X}/L_{X}= v_{rec}/v_A$) and equation (\ref{eq:vrec_turb}) imply that

\begin{equation}
  \Delta R_{X} = L_{X}  \min\left[{L_{inj}\over L_X}, {L_X\over L_{inj}}\right]^{1\over2}M_{A}^2.
\end{equation}
 
We may employ this relation to compute the magnetic reconnection power from eq.(\ref{eq:wb00}). Considering the discussion above, we may assume that the injection velocity of the turbulence is of the order of the coronal gas sound speed and the injection scale of the turbulence is of the order of the size of the reconnection zone, i.e., $L_{inj} \simeq L_X$, which in turn may be set as a free parameter in the model being smaller than the coronal loop size, i.e., $L_X \leq L$. With these assumptions, we may rewrite the equation above as

% equation ok!
\begin{equation}
  \Delta R_{X} \simeq 2.34 \times 10^4 \Gamma^{-{5\over16}} r_{X}^{31\over64} l^{-{5\over32}} l_{X}q^{-{3\over4}} \dot{m}^{-{5\over32}} m~~cm,
\label{eq:deltar_turb}
\end{equation}
where $l_{X}=L_{X}/R_{S}$.

This results a magnetic reconnection width which is much larger than in the case of anomalous resistivity (eq.\ref{eq:deltar}) indicating that reconnection driven by turbulence can be much more efficient to drive fast reconnection. As a matter of fact, observations of solar flares indicate fast reconnection rates up to $v_{rec}/v_A \sim 0.1$ \citep{takasaki_etal_04}, while numerical simulations of turbulent fast reconnection also result values up to $v_{rec}/v_A \sim 0.1$ depending on the turbulent injection power ($P \propto v_{inj}^2$). Here, because of the lack of knowledge of the physical details of the injection and power scales of the turbulence, we adopt the analytical relation above for $\Delta R_{X}$. 

The fact that turbulence results a much larger reconnection rate than anomalous resistivity is actually not a surprise and comes directly from the different nature of both processes. As stressed before, anomalous resistivity acts only at small scales resulting a much smaller reconnection rate, while collisional turbulence acts on the large scales of the fluid, as we can see, for instance, by comparing the equation above with eq. (\ref{eq:deltar}).
\citet{kowal_etal_09} have also compared the two processes employing 3D MHD numerical simulations and found that the introduction of anomalous resistivity is unable to affect the rate of reconnection due to  weak turbulence. 
This  supports the notion that in the presence of turbulence and magnetic field stochasticity induced by it, plasma kinetic effects do not seem dominant in the determination of the global reconnection speed.

This point can be better understood considering the condition for having turbulent collisional reconnection according to the LV99 theory. This requires the thickness of the reconnection region (eq.\ref{eq:deltar_turb}) to be larger than the ion Larmor radius, $\Delta R_X > R_i$ \citep{eyink_etal_11}.
We find that this condition  is satisfied in all the fiducial parametric space investigated here for the physical conditions around the BHs.
On the opposite situation, collisionless kinetic effects would prevail to induce fast reconnection.
 
The result above also suggests that anomalous resistivity may be an important process in the beginning of the reconnection process, but once turbulence is developed in the system, then fast reconnection induced by turbulence will be the dominant process.

Considering the equations (\ref{eq:wb00}) and (\ref{eq:deltar_turb}), we obtain that the magnetic reconnection power released by turbulent fast reconnection in the surrounds of the BH is given by

%equation ok!
\begin{equation}
 \dot{W}_B \simeq 1.66\times10^{35} \Gamma^{-{1\over2}} r_{X}^{-{5\over8}} l^{-{1\over4}} l_{X} q^{-2} \dot{m}^{3\over4} m~~erg/s,
 \label{eq:Wbturb}
\end{equation}
which obviously results a larger value than in the case of fast reconnection driven by anomalous resistivity (eq.\ref{eq:Wb}).

\begin{figure}[t]
 \centering
 \includegraphics[width=0.5\textwidth] {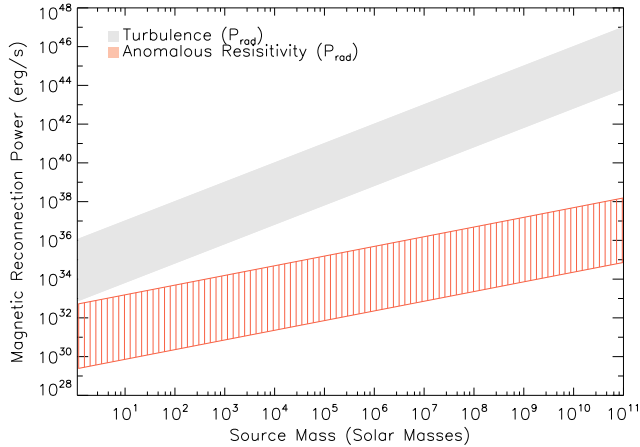}
 \caption{Magnetic power ($\dot{W}_{B}$) released by fast reconnection driven by anomalous resistivity  (in red) and by turbulence (in light gray) as a function of the BH mass. The parametric space spans $0.05 \leq \dot{m} \leq 1$; $1 \leq l \lesssim 18$; and $0.06 l \lesssim l_X \leq l$ (assuming $R_{X}=6R_{S}$).}
 \label{fig:turbares}
\end{figure}

Figure \ref{fig:turbares} compares the two values derived for the magnetic reconnection power (equations \ref{eq:Wb} and \ref{eq:Wbturb}) as a function of the mass of the central source for a suitable choice of parameters: $R_{X}=6R_{S}$; $1 \leq m \leq 10^{10}$ (in $M_{\odot}$ units) to spam masses from microquasars to AGNs; and $0.05 \leq \dot{m} \leq 1$ (in $\dot{M}_{Edd}$ units) for the mass accretion rate. Also, in  order to ensure near collisionality in the flow and the validity of the equations above in the parametric space, we have constrained the lower bound of the characteristic scales of the system to be larger than the ion mean free path. Specifically, we have imposed $l_{mfp} \lesssim l_{X} \leq l$, where $l_{mfp} \simeq 5.70\times10^{-2} l~cm$ is the mean free path in $R_{S}$ units. 
This gives  $1 \leq l \lesssim 18$ (in $R_{S}$ units) for the length of the magnetic loop (or the height of the corona) and $0.06 l \lesssim l_X \leq l$ (in $R_{S}$ units). The upper limit of $l$ has been obtained from the condition $l_{mfp} \lesssim 1 $.

While in the anomalous resistivity case $\dot{W}_B$ has a dependence with the source mass given by $\dot{W}_{B} \propto m^{0.5}$ in the turbulent reconnection case this dependence is steeper $\dot{W}_{B} \propto m$ as evidenced in the Figure \ref{fig:turbares}. We will see in Section \ref{sec:comparison} that this has important observational consequences for microquasars and AGNs.

\subsection{Radiation-pressure dominated versus gas pressure-dominated accretion disks}
\label{sec:radgas_regime}

The values of accretion rate employed in Figure \ref{fig:turbares} are more suitable for a corona connected to a radiation-pressure dominated disk. We may also evaluate $\dot{W}_B$ from equations (\ref{eq:Wb}) and (\ref{eq:Wbturb}) when considering a gas-pressure dominated disk. In this case, the accretion rates cannot be as large as those considered in the radiation regime in Figure \ref{fig:turbares}.
 
In a gas-pressure dominated regime the disk pressure is given by \citet{shakura_sunyaev_73}:

% equation ok!
\begin{equation}
  P_{gas} \simeq 4.11 \times 10^{18} \alpha^{-{9\over10}} r_{X}^{-{21\over8}} q^{17\over5} \dot{m}^{17\over20} m^{-{9\over20}}~~dyn/cm^2,
\label{eq:pgas}  
\end{equation}
while for a radiation pressure dominated regime the disk pressure is:
% equation ok!
\begin{equation}
  P_{rad} \simeq 4.78 \times 10^{15} \alpha^{-1} r_{X}^{-{3\over2}} m^{-1}~~dyn/cm^2,
\label{eq:prad}
\end{equation}
where $\alpha$ is the disk viscosity parameter. 

We can compare the two equations above in order to obtain the range of accretion rates which are suitable for each regime. 
Figure \ref{fig:pp} in Appendix \ref{app:prad_pgas} depicts the ratio between these two pressures as a function of the mass of the sources. We see that for $0.05 \leq \alpha \leq 0.5$ \citep[see][]{King_etal07}, considering the whole range of masses, $P_{rad}/P_{gas} < 1$ (gas-pressure dominated regime) for $\dot{m} \leq 5\times10^{-4}$, and $P_{rad}/P_{gas} > 1$ (radiation-pressure dominated regime) for $\dot{m} > 5\times10^{-2}$. In the next sections, we will adopt these ranges of accretion rates in the computation of the magnetic reconnection power for each accretion disk regime.

\section{Comparison of $\dot{W}_{B}$ with the observed core radio and gamma emission of microquasars and AGNs}
\label{sec:comparison}

In the previous section, we evaluated the magnetic reconnection power produced by fast reconnection in the surrounds of a BH, considering two different mechanisms to induce it, anomalous resistivity and turbulence. The second one was found to be much more efficient. It is out of the scope of the present work to predict what amount of this magnetic power goes to accelerate particles, but as stressed in section \ref{sec:intro} (also in GL05 and GPK10), we may expect that a substantial fraction of it will produce high-speed electrons which will spew outward \citep{kowal_etal_11,kowal_etal_12} and produce relativistic ejecta and Synchrotron radio emission. In this section, we first compare the calculated fast magnetic reconnection power with the observed radio emission of the nuclear regions of microquasars and AGNs, then we compare this power with the processed VHE emission from these sources.

\subsection{Low-luminosity sources}

\begin{figure}[t]
 \centering
 \includegraphics[width=0.5\textwidth] {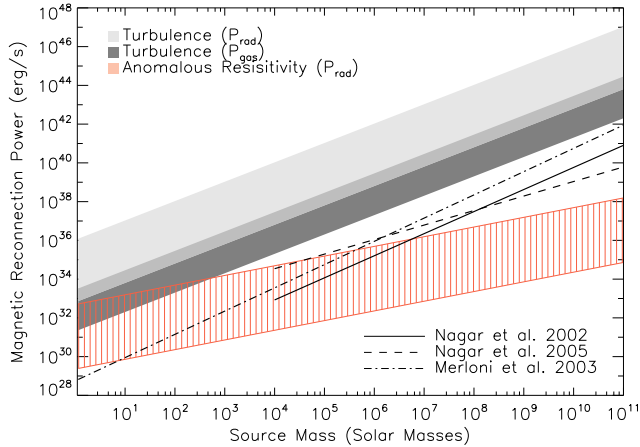}
 \caption{Magnetic power ($\dot{W}_{B}$) released by fast reconnection driven by anomalous resistivity (in red) and by turbulence as a function of the BH mass. The upper part of the diagram of the turbulent driven reconnection power (light gray) corresponds to a radiation-pressure dominated disk with larger accretion rates ($0.05 \leq \dot{m} \leq 1$), while the lower part of the diagram (dark gray) stands for a gas-pressure dominated disk  with $\dot{m} \simeq 5\times10^{-4}$ (see appendix \ref{app:prad_pgas} and Figure \ref{fig:pp} for details); and the intermediate gray region is the overlap between both regimes. 
The diagram of the  magnetic reconnection power driven by anomalous resistivity also corresponds to a radiation-pressure dominated disk. The other free parameters in the diagrams span $1 \leq l \lesssim 18$; and $0.06 l \lesssim l_X \leq l$ (assuming $R_{X}=6R_{S}$), as in Figure \ref{fig:turbares}. The continuous and dashed lines correspond to the observed correlations between the BH mass and the core radio luminosity found for LLAGNs by \citet{nagar_etal_02} and \citet{nagar_etal_05}, respectively; and the dot-dashed line corresponds to observed correlations for AGNs and microquasars by \citet{merloni_etal_03}.}
 \label{fig:radio01}
\end{figure}

Figure \ref{fig:radio01} depicts in gray color scales the calculated magnetic power released by a fast magnetic reconnection event as a function of the central BH mass induced both by turbulence (eq.\ref{eq:Wbturb}) in radiation and gas-pressure dominated regimes and by anomalous resistivity in a radiation pressure regime (eq.\ref{eq:Wb}). The continuous line in the figure corresponds to the observed correlation between the BH mass and the core radio luminosity for a sample of $96$ nearby LLAGNs (within distances of $19$ Mpc) found by \citet{nagar_etal_02} from VLA and VLBA observations. The dashed line was obtained by \citet{nagar_etal_05} considering a more refined VLBI sample.
 The dot-dashed line corresponds to the observed correlations by \citet{merloni_etal_03} considering the VLA  $5$GHz core radio emission of a sample of $\sim 100$ AGNs (most of which with arcsecond resolution) and the radio emission of $8$ galactic black holes obtained with the Green-Bank Interferometer. Despite the simplicity of our model, the slope dependence of the magnetic power released by turbulent reconnection with the source mass is very similar to the observed radio luminosity-source mass correlations for these sources. A closer examination of the diagram shows that the predicted intensities of the turbulent driven magnetic reconnection power in the swept parametric space are much larger than the observed radio luminosities, specially in the upper part of the diagram, i.e., in the radiation pressure-dominated case. This is an indication that in general only a small fraction of the magnetic reconnection power would be enough to explain the observed radio emission for most of the sources represented by the correlation lines. The anomalous resistivity model on the other hand does not match most of the observed correlations for the parametric space considered. In GPK10 work, the comparison of the magnetic reconnection power driven by anomalous resistivity with the radio luminosity of a bunch of microquasars and LLAGNs had revealed a better match between both because in that case, the magnetic reconnection power was computed considering a larger disk radiation field than the present evaluation (eq.\ref{eq:urad}) making the corona hotter (eq.\ref{eq:tcr02}) and therefore, the magnetic reconnection more efficient. Also, in GPK10 we considered coronal heights up to nearly $1000 R_S$ which in that approach made the upper bound of the reconnection power larger than in the present case. 

\begin{figure*}[t]
 \centering
 \includegraphics[width=0.8\textwidth]{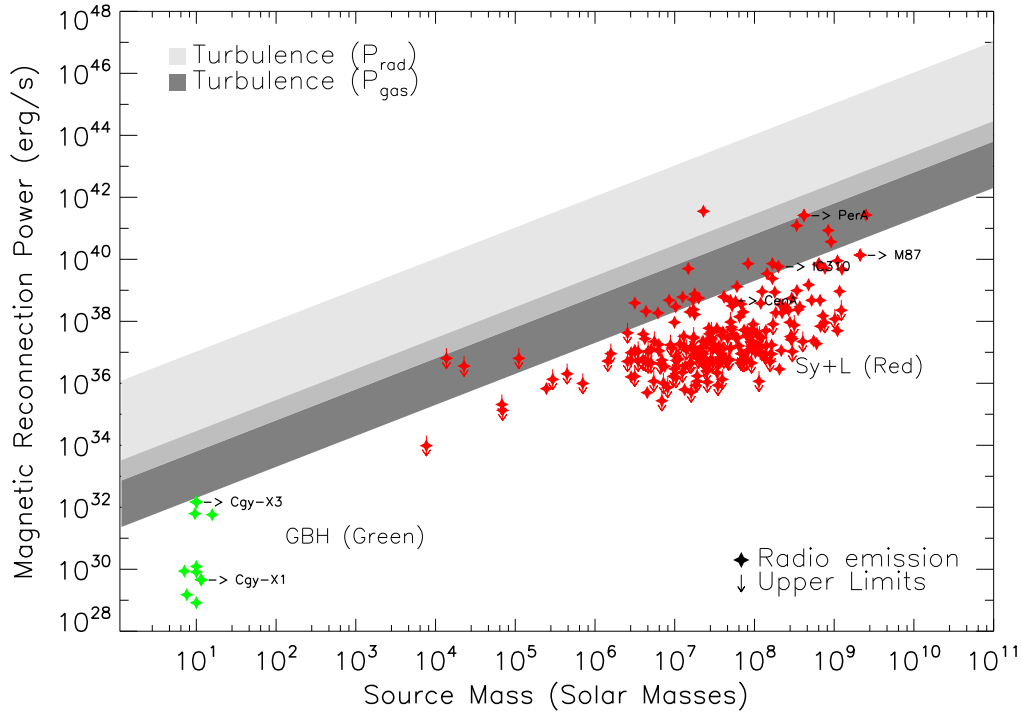}
 \caption{Turbulent driven magnetic reconnection power $\dot{W}_{B}$ (eq.\ref{eq:Wbturb}) against BH source mass, as in Figure \ref{fig:radio01}, compared to the observed core radio emission of $233$ LLAGNs and $9$ microquasars (GBHs). The parametric space used in the calculation of $\dot{W}_{B}$
 is the same as in  Figure \ref{fig:radio01}. 
 The green diamonds give the observed core radio luminosities for microquasars, the red diamonds the core radio luminosities of LLAGNs (LINERS and Seyfert), and the arrows indicate upper limits data. The position of the observed emission for a few sources is highlighted with black arrows.}
 \label{fig:graf_radio}
\end{figure*}

Figure \ref{fig:graf_radio} compares the calculated magnetic reconnection power driven by turbulence with the observed core radio luminosities of a large sample of LLAGNs and microquasars, namely $9$ microquasars \citep[or galactic black holes - GBHs, see][]{hannikainen_etal_01, merloni_etal_03}, and $233$ LLAGNs \citep[including Seyferts and LINERs galaxies, see][]{merloni_etal_03, nagar_etal_02, nagar_etal_05, israel_98, kadler_etal_12}.  Table \ref{tab:radio} lists the physical parameters for this sample. 
 The BH masses of the sources (also indicated in  Table \ref{tab:radio}) were evaluated from averages taken from different determinations in the literature including
kinematic methods using stellar, gas, or maser dynamics \citep[][]{richstone_etal_98, gebhardt_etal_00, merloni_etal_03, remillard_mcclintock_06}, and
the relation between the BH mass and the dispersion velocity at the center of the host galaxy \citep{tremaine_etal_02,merritt_ferrarese_01} obtained from the HYPERLEDA catalogue\footnote{http://www-obs.univ-lyon1.fr/hypercat/}.
 
The radio emission from the sources in Figure \ref{fig:graf_radio} is represented by diamond symbols, with the red color corresponding to LLAGNs (LINERs an Seyfert galaxies) and the green color to microquasars. We also  highlighted the location of the observed radio emissions of a few sources that have been extensively explored in multi-wavelength campaigns, i.e., the radio galaxies Cen A, M$87$, IC $310$, and Per A (NGC $1275$), and the microquasars Cgy-X$1$ and Cgy-X$3$. Figure \ref{fig:graf_radio} confirms the trend of the previous Figure \ref{fig:radio01} indicating that the magnetic reconnection power extracted from reconnection of the magnetic lines in the inner coronal region around the BHs of microquasars and LLAGNs could be enough to explain the core Synchrotron radio emission from them, as suggested already by GPK10 for a much smaller sample of sources (and considering a fast magnetic reconnection model driven by anomalous viscosity). Actually, the results indicate that for most of the sources only a small fraction of the magnetic reconnection power would be sufficient to accelerate the electrons responsible for the radio Synchrotron emission. 

\begin{figure*}[t]
 \centering
 \includegraphics[width=0.8\textwidth] {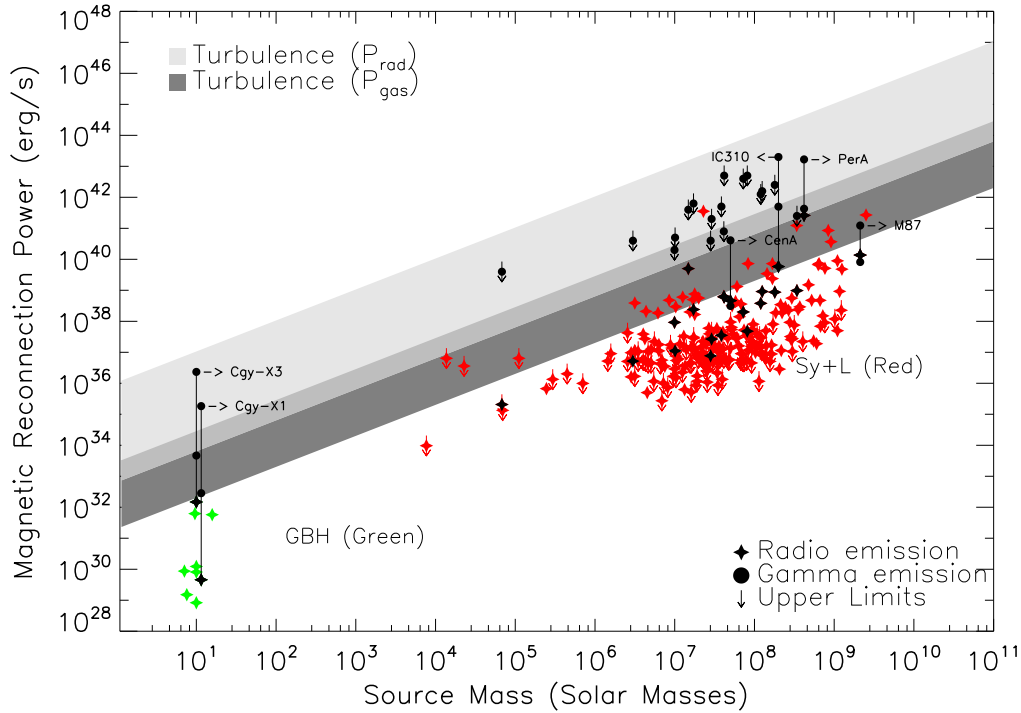}
 \caption{The same as in Figure \ref{fig:graf_radio}, but now including also the gamma-ray emission of a subsample of LLAGNs, and microquasars for which this emission is available. This emission is represented by black circles.  The radio emission of the corresponding sources is also represented by black stars to distinguish them from the rest of the sample. In a few cases for which there is observed gamma-ray luminosity from  MeV/GeV to TeV ranges, we plotted the maximum and minimum values linking both circles with a vertical black line that extends down to the radio emission of each source. The arrows associated to some sources indicate that the gamma-ray emission is an upper limit only.
}
 \label{fig:radio_gamma00}
\end{figure*}

Laboratory experiments of magnetic reconnection \citep[][]{yamada_etal_14} and solar flare observations \citep{lin_hudson_71} indicate that $\sim 50-60\%$  of the  magnetic power released by reconnection can go into particle acceleration. Figure \ref{fig:radio01} indicates that there is in general much more power available to accelerate particles than that spent in the radio Synchrotron emission.
As discussed in Section \ref{sec:intro}, accelerated relativistic electrons along with accelerated protons will also cool via other mechanisms that will lead to HE and VHE emission. These processes include inverse Compton (IC) relativistic electron interactions with the surrounding photon field, or Synchrotron-self Compton (SSC) interactions with the Synchrotron photons they produce, or proton interactions with surrounding protons (via p-p interactions) and photons (via p-photon interactions) \citep[e.g.,][and references therein]{romero_etal_03, khiali_etal_14, aleksic_etal_10b, abdo_etal_09a,abdo_etal_09b}.
For this reason in Figure \ref{fig:radio_gamma00} we have also plotted the observed gamma-ray luminosities (see Table \ref{tab:radio}) which is available for only a subsample of $23$ sources of those depicted in Figure \ref{fig:graf_radio}. For most of the Seyferts galaxies of this subsample, the figure shows only upper limits of the gamma-ray luminosity in the GeV band (obtained with $95\%$ confidence level
by \citet[][]{ackermann_etal_12} with Fermi-LAT). These data are represented by circle symbols in Figure \ref{fig:radio_gamma00}, and Table \ref{tab:radio} provides more information on the gamma-ray luminosity for these sources. We included also the observed gamma-ray luminosities of the four radio galaxies highlighted in the figure \citep{abdo_etal_09b, abdo_etal_10, aleksic_etal_14a,aleksic_etal_14b}, as well as of the microquasars Cgy-X$1$ \citep[][]{albert_etal_07,malyshev_etal_13} and Cgy-X$3$ \citep[][]{piano_etal_12}. In these cases, there is data available from MeV/GeV to TeV bands and we represented the whole luminosity range (from the maximum to the minimum observed values) with circles linked by vertical lines that also extend  to the radio emission of each of these sources (see also Table \ref{tab:radio}).

Figure \ref{fig:radio_gamma00} shows that the magnetic reconnection power diagram 
also encompasses the observed gamma-ray emission of the low luminosity sources, though we cannot predict what fraction of this power might be required to produce it.
In fact, because the HE and VHE emission of the sources do not depend only on the energy of the accelerated primary relativistic particles, but also on the surrounding photon and proton density fields of the source, one should not expect that the magnetic reconnection power might also directly probe this emission. If the released $\dot{W}_{B}$ is the responsible for the acceleration of the relativistic particles then, it must be larger than (or at least comparable, depending on the energy transfer efficiency) to the associated electron radio synchrotron radiation, as we find.   
 Nevertheless, a striking feature in Figure \ref{fig:radio_gamma00} is the fact that there is in principle power enough to produce also the gamma-ray emission (of course, with different amounts for different sources) and this emission nearly follows the same trend of the observed radio emission in the diagram, and thus both emissions are correlated and could be possibly produced in the same region in the core, by the same relativistic particle populations, which in turn can be due to the magnetic reconnection mechanism as described.
 
 We note that upper limits of the gamma-ray luminosity of a much larger sample of Seyfert galaxies obtained by \citet[][]{ackermann_etal_12}, but with no counterpart in the radio emission sample of Figure \ref{fig:radio_gamma00}, were also found to match and follow the same trend of the other sources represented in the diagram of Figure \ref{fig:radio_gamma00}.
  
\subsection{Inclusion of high-luminosity sources}
\label{section:HLAGNs}

In the previous section, we discussed the correlation between the magnetic power released by fast magnetic reconnection at the inner corona/accretion disk region, and the observed radio and gamma-ray luminosities of LLAGNs and microquasars. In this section we
extend this analysis to a much broader sample that includes blazars and GRBs, i.e., high-luminosity sources.

\begin{figure*}[t]
 \centering
 \includegraphics[width=0.8\textwidth] {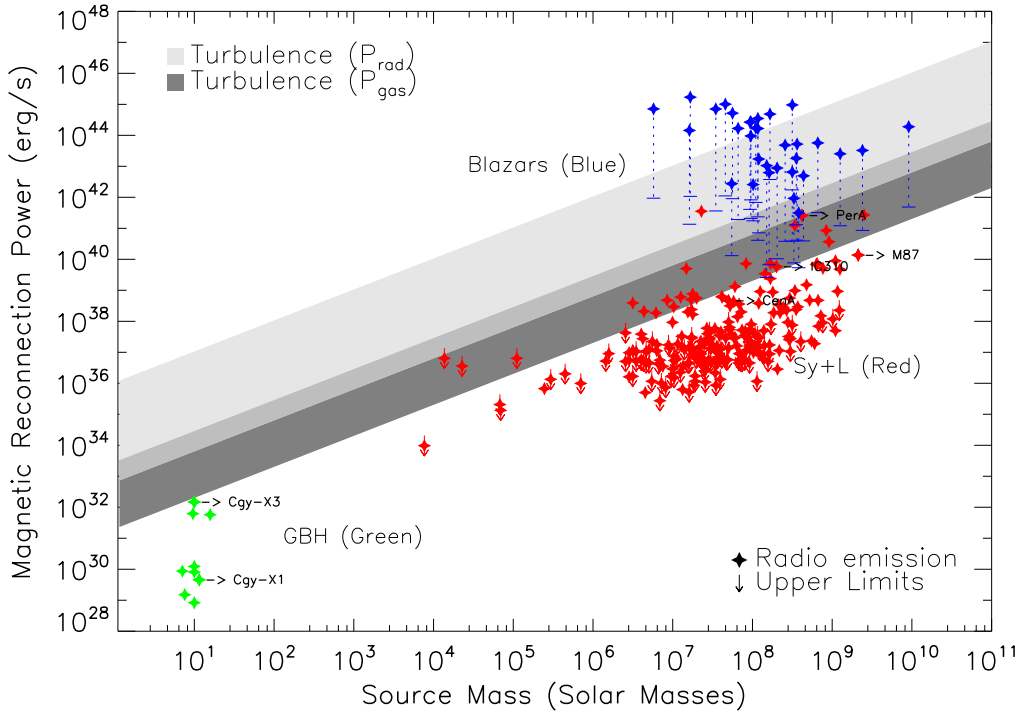}
 \caption{The same as in Figure \ref{fig:graf_radio}, but now including also the observed radio emission from blazars (represented in blue diamonds). The vertical dashed lines correct the observed emission by the effects of Doppler boosting (see more details in the text).}
 \label{fig:radio_radio}
\end{figure*}

  As stressed in Section \ref{sec:intro}, in blazars the jet is known to point towards the line of sight screening most of the inner core radiation, but the observed radio emission is often separated in a core (probably produced near the jet basis) and an extended component \citep[e.g.,][]{kharb_etal_10}. Figure \ref{fig:radio_radio} and  Table \ref{tab:blazar} present the core (or jet basis) radio emission of these blazars along with the core radio emission of LLAGNs.  As in Figure \ref{fig:graf_radio}, we compare the observed emission of these sources with the calculated turbulent driven magnetic reconnection power. 
 This sample has $32$ blazars studied by \citet{nemmen_etal_12} (whose black hole masses and radio emission were obtained from \citet{vovk_neronov_13} and \citet{kharb_etal_10}, respectively). The dashed vertical bars associated to the blazars emission give the corrected radio luminosity due to Doppler beaming. We performed the same correction as \citet{nemmen_etal_12} did for the gamma-ray luminosity of these sources.\footnote{\citet{nemmen_etal_12} assumed an isotropic gamma-ray emission  and then  corrected it by the beaming factor $f_b$, i.e., $L_{corr}=f_{b}L_{iso}$, where $L_{corr}$ is the corrected luminosity, $L_{iso}$ is the isotropic luminosity, and $f_{b}=1-\cos(1/\Gamma_{0})$, where $\Gamma_{0}$ is the bulk Lorentz factor.} 

Figure \ref{fig:radio_radio} indicates two striking features, with the Doppler boosting correction, most of the blazars radio emission lies in the highest accretion rate part of the $\dot{W}_{B}$ diagram. This is compatible with the notion that this emission is actually produced near the jet launching basis and therefore, could well be triggered by fast magnetic reconnection as in the LLAGNs. On the other hand, in spite of the uncertainties in the determination of the Doppler boosting correction and the poorness of the sample, the blazar emission does not seem to follow the same trend as that of the LLAGNs, specially the highest luminosity ones. This may be an indication that their emission, specially that of the highest luminosity sources, comes mostly from further out in the jet basis, and may be due to another population of relativistic particles, probably produced already at the jet shock, as it is generally expected. 
 
\begin{figure*}[t]
 \centering
 \includegraphics[width=0.8\textwidth] {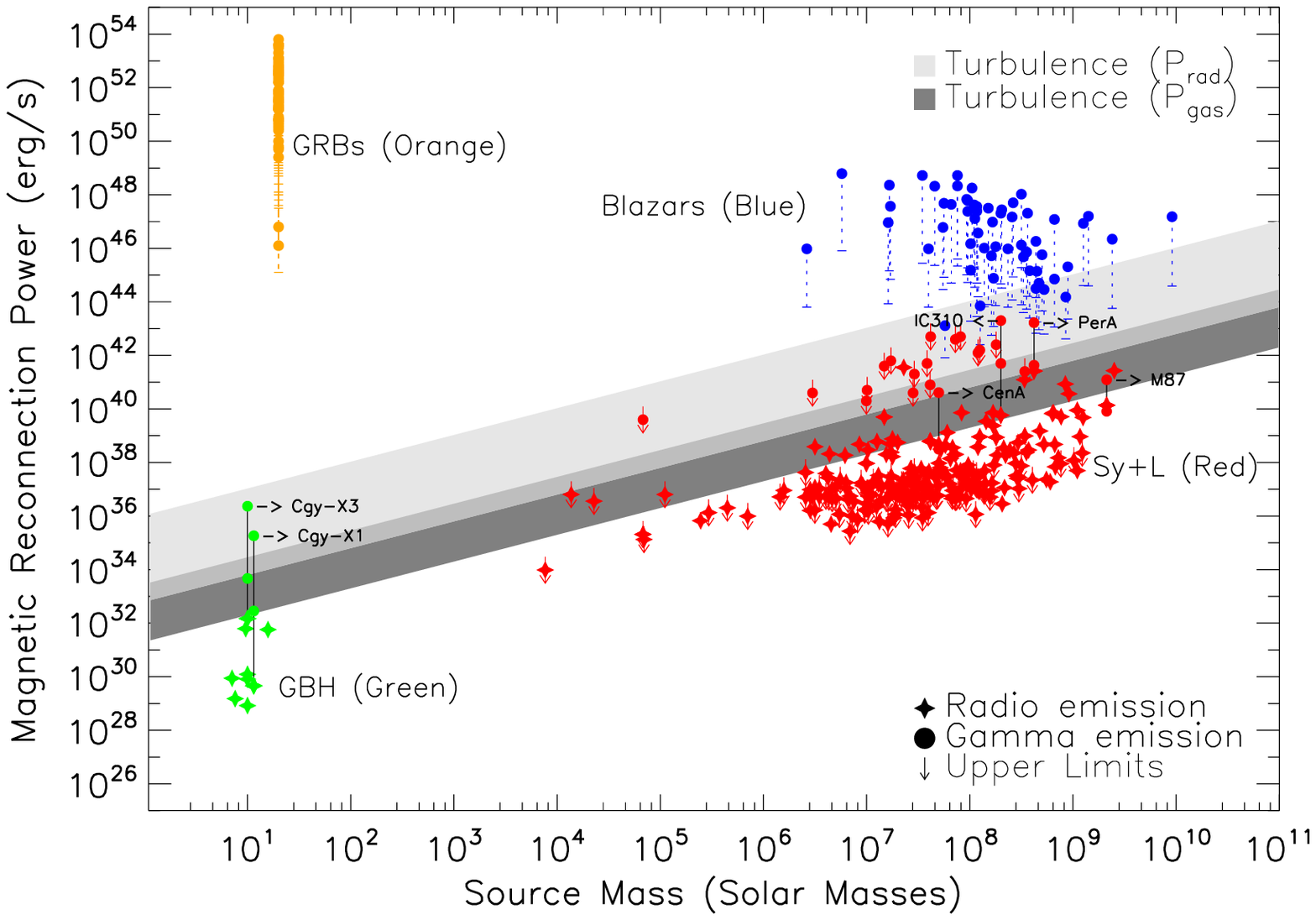}
 \caption{The same as in Figure \ref{fig:radio_gamma00}, but now including also the gamma-ray emission from blazars (represented in blue circles) and GRBs (represented in orange circles). The vertical dashed lines correct the observed emission by the effects of Doppler boosting (see more details in the text).}
 \label{fig:radio_gamma}
\end{figure*}

To strengthen the conclusion above, in Figure \ref{fig:radio_gamma} we have plotted the observed gamma-ray emission  for the $32$ blazars \citep[][see also Table \ref{tab:blazar}]{nemmen_etal_12} along with the observed gamma and radio emission of the LLAGNs. Even with the Doppler correction, we note that most of the blazars gamma-ray emission lies above the $\dot{W}_{B}$ diagram. Besides, as the radio emission, it does not follow the same trend of the LLAGNs. 

For comparison, we have also plotted the gamma-ray emission of a sample of $54$ GRBs \citep[also obtained from][see Table \ref{tab:grb}]{nemmen_etal_12}.
Figure \ref{fig:radio_gamma} clearly shows that neither the blazars nor GRBs have their gamma-ray luminosity  correlated with $\dot{W}_{B}$. Actually, the trend that links these two classes of sources in the diagram is anti-correlated with the gray zone that characterizes the nuclear magnetic reconnection emission mechanism here discussed. This suggests that the gamma emission in these sources cannot be attributed to this emission process. This result is consistent to what is usually expected in the case of the GRBs, i.e., that their gamma-ray emission does not come from the core or the engine which is totally embedded in the envelope of the progenitor \citep[e.g.,][]{paczynski_98, macFadyen_woosley_99}. In the case of the blazars, this result also indicates that their nuclear emission is probably obscured by the jet pointing to us and therefore, what we really observe is gamma-ray emission coming predominantly  from the optically thin jet.  
 
\section{Discussion and Conclusions}
\label{sec:conclusions}

We have here extended the earlier work by GL05 and GPK10, investigating the interactions between the magnetic field lines that arise from the accretion disk and the magnetosphere anchored into the BH horizon of  microquasars and AGNs. We examined the conditions under which fast magnetic reconnection events can occur in this inner region and release enough magnetic power to enable the heating and acceleration of particles to relativistic velocities. 

\subsection{Anamalous resistivity versus turbulence induced fast reconnection and other processes}
Reviewing the potential mechanism(s) that can excite fast reconnection in this nearly collisional MHD scenario, we compared the earlier investigated process in GL05 and GPK10, namely, anomalous resistivity \citep[][]{parker_79, biskamp_etal_97, shay_etal_98}, with reconnection driven by turbulence \citep[][LV99]{kowal_etal_09}. 
We have found that the magnetic power released by turbulent driven fast reconnection (eq.\ref{eq:Wbturb}) is much larger  than that obtained by anomalous resistivity (eq.\ref{eq:Wb}), as the presence of turbulence naturally enlarges the thickness and the volume of the magnetic reconnection discontinuity (eq.\ref{eq:deltar_turb}).  
This result is not a surprise, since 
anomalous resistivity acts dominantly at small scales resulting a much smaller reconnection rate, while collisional turbulence acts on the large scales of the fluid (see eqs. \ref{eq:deltar} and \ref{eq:deltar_turb} that compare the thickness of the reconnection zone for both processes).

Back to 2005, GL05 (see also GPK10) explored only the effects of anomalous resistivity in driving fast reconnection in the surrounds of BHs, because this process was already largely studied, while the LV99 theory was still under testing. Currently, LV99  theory has been thoroughly discussed considering different approaches \citep[e.g.,][]{eyink_etal_11, lazarian12, lazarian15} and  successfully tested by means of 3D MHD simulations \citep[][]{kowal_etal_09, kowal_etal_12,eyink_etal_13, xu13}, therefore motivating its examination in the present analysis. Interestingly, we have found that  it is able to reproduce much better the observations (see section \ref{sec:comparison} and below).  

The perception that turbulence might affect magnetic reconnection (as in LV99) is not unprecedented \citep[for a review with a comparative analysis of the different models see][]{kowal_etal_09, eyink_etal_11, lazarian15}. Several earlier works focussed on the effects of turbulence  at  microphysical scales \citep[e.g.,][]{speiser_70,jacobson_moses_84}, but at the MHD large scale level these kinetic effects are not dominant. The closest study to LV99 model was done by \citet{matthaeus_lamkin_85,matthaeus_lamkin_86}. These authors explored 2D magnetic reconnection in the presence of  turbulence and found a significant enhancement in the reconnection rate. However, they did not derive an analytical prediction for the reconnection speed. Other works have introduced the hyper-resistivity concept and tried to derive fast reconnection rates from turbulence within the context of mean-field resistive MHD \citep{strauss_86, bhattacharjee_hameiri_86, hameiri_bhattacharjee_87, diamond_malkov_03, yokoi_hoshino_11, guo_etal_12}. Though the approach at first level seems interesting, these works have reached different results in the estimates of the reconnection rate and besides, they still lack multidimensional numerical testing.

Several possibilities  of fast reconnection in the collisional MHD regime driven by instabilities in  the reconnection layer have been also largely discussed \citep[e.g.,][]{loureiro_etal_09, bhattacharjee_etal_09}. As remarked in \citet{eyink_etal_11}, these ideas can be traced back to the work of \citet{shibata_tanuma_01} \citep[see also][]{strauss_86, waelbroeck_89} who suggested that tearing instability may result in fractal reconnection taking place on very small scales. Estimates indicate that laminar current sheets subject to tearing instability have reconnection rates that are a little faster than the Sweet-Parker, but they  enlarge the reconnection layer enabling a wide outflow which will become turbulent in most astrophysical conditions. In this case, the instability can be important for initiating reconnection when the level of pre-existing turbulence is still low, but once turbulence becomes dominant  this will dominate reconnection making it very fast. 
In conclusion, like anomalous resistivity, we expect these instabilities to be important for the onset of reconnection and  turbulence, therefore increasing the three-dimensional stochasticity of magnetic field lines and thus initiating large scale fast reconnection, as proposed in LV99.

\subsection{Particle acceleration induced by the magnetic power released by reconnection} 
In the present work we focussed on the derivation of the magnetic power released by fast magnetic reconnection and then, arguing  that part of this energy would be able to accelerate particles to relativistic velocities,  we compared this power  with  the observed radio and gamma-ray luminosities of BH sources spanning $10^{10}$ orders of magnitude in mass and $10^6$ orders of magnitude in luminosity. We found   that these luminosities and therefore, the relativistic particle population responsible for them could be due to this magnetic power (see Section 4.3).

Though the specific nature of the particle acceleration mechanism is not a critical point in the present work, a few notes are in order in this regard. Particle acceleration by a first-order Fermi process in fast magnetic reconnection sites has been extensively studied \citep[see, e.g., the reviews in][and references therein]{kowal_etal_11,kowal_etal_12,dgdp_etal_13, dgdp_etal_14}. As remarked before, GL05 were the first to propose that this process might occur within current sheets. In analogy to shock acceleration, GL05 verified that  particles  trapped between the two converging magnetic fluxes (moving to each other with a velocity $V_R$), would  bounce back and forth undergoing head-on interactions with magnetic fluctuations and their energy after a round trip would increase by $<\Delta E/E> \sim V_R / c$, which implies a first-order Fermi process with an exponential energy growth after several round trips,  resulting a power-law particle spectrum.
Before that, several authors \citep[e.g.,][]{litvinenko_96, shibata_tanuma_01, zenitani_hoshino_01} addressed the acceleration of energetic particles in reconnection sites but did not describe the process as a first-order Fermi process. The analytical predictions of GL05 were confirmed by  \citet{drake_etal_06} who made 2D numerical PIC simulations of the process in collisionless fluids which thus work only for 2D converging magnetic islands and probe only kinetic scales \citep[see also,][]{zenitani_hoshino_08, zenitani_etal_09}. The equivalence between the two models was discussed in \citet{kowal_etal_11,kowal_etal_12} who performed 2D and 3D numerical collisional MHD simulations with test particles. Besides, these authors demonstrated that the process works in 3D fluids (where the 2D magnetic islands are opened into 3D loops, as described in GL05).
 
 These results strengthen the possibility that the overall magnetic reconnection process in the surrounds of a BH, near the jet launching basis, can lead to particle acceleration and allow for the observed  Synchrotron radio emission in the core regions of these sources and the associated high energy gamma-ray emission as well\footnote{Of course, as stressed before, we cannot exclude the possibility that fast magnetic reconnection may also lead to the production of plasmoids that  can  propitiate the formation of a shock right outside of the reconnection region allowing for  particle acceleration in this shock \citep[see also][]{khiali_etal_14}.}. 

\subsection{Comparison with observations}
Derived as a function of inner radius region parameters, i.e., the mass of the central BH ($m=M/M_{\odot}$), the disk mass accretion rate ($\dot{m}=\dot{M}/\dot{M}_{Edd}$), the extension of the coronal loops ($l = L/R_{S}$), and the extension of the magnetic reconnection region in the corona ($l_{X}=L_{X}/R_{S}$),
 the calculated fast magnetic reconnection power driven by turbulence was compared with the observed nuclear radio and gamma-ray emission of a much larger sample of compact sources than that used in GPK10, including microquasars, low-luminosity AGNs (LLAGNs, i.e., LINERS and Seyfert galaxies), as well as blazars and GRBs. 
Our results show that, in general, just a small fraction of this power would be enough to explain the observed radio luminosities of the low-luminosity sources (LLAGNs and microquasars) (see Figure \ref{fig:graf_radio}). In most of these cases, the corresponding required accretion rate is $\dot{m}<0.05$.  Also striking is the fact that the slope dependence of the magnetic power released by turbulent reconnection with the source mass nearly follows the same trend of the observed radio luminosity-source mass correlation for these sources \citep[][see Figure \ref{fig:radio01}]{nagar_etal_02,nagar_etal_05, merloni_etal_03}, which suggests that this mechanism could provide a natural interpretation for this correlation, as suggested earlier by GPK10, but considering a very small sample of sources and fast reconnection induced by anomalous resistivity only. 

The corresponding gamma-ray emission of these sources, which is believed to be produced by a number of leptonic and hadronic processes involving the accelerated electrons and protons, respectively (through synchrotron-self Compton, inverse Compton, proton-proton inelastic collisions, and proton-photon collisions \citep[][]{romero_etal_03, khiali_etal_14}, can in principle be also associated with the same emission zone that produces the radio synchrotron emission in the core of these sources. 
For this reason, we have investigated whether the power released by magnetic reconnection could also  be connected with the gamma-ray emission of these sources. We see that this could  be the case for microquasars and LLAGNs.  
The observed gamma-ray luminosity of these sources is nearly correlated with both the radio luminosity and the calculated magnetic reconnection power (Figure \ref{fig:radio_gamma00}), being smaller than the latter. This suggests that the accelerated particles by the magnetic reconnection mechanism here discussed can produce the radio emission and be also responsible for the processing of the high energy emission in the core region. Even the radio galaxy IC $310$ which has been argued to emit like a blazar \citep[e.g.,][]{aleksic_etal_14a}, follows this trend.
We note however that, because in most cases the observed gamma-ray luminosity is larger than the radio luminosity, the former lies, in general, in the upper parts of the magnetic reconnection diagram, therefore corresponding to accretion rates  which are larger than those predicted when examining only the radio emission of the sources. This suggests that in most cases, in order to produce magnetic power enough to accelerate particles able to produce both the radio and the VHE  emission, the accretion rates must be in general $\dot{m} \geq 0.05$.

Moreover, the correlations found above may also shed some light in the interpretation of the so-called ``fundamental plane'' obtained empirically, which correlates the radio and X-ray emission of microquasars and low-luminosity AGNs with the BH mass of the sources \citep[see][]{merloni_etal_03, falcke_etal_04, wang_etal_06, kording_etal_06, li_etal_08, yuan_etal_09, gultekin_etal_09, plotkin_etal_12,huang_etal_14}. Although we here did not deal with the X-ray emission which is directly related with the accretion disk processes, but focussed on the radio and VHE emissions related to the disk corona, our model suggests a simple physical interpretation for the existence of these empirical correlations as linked to magnetic reconnection activity in the core of these sources. The fact that fast magnetic reconnection and the associated radio flare is a transient and violent phenomenon that leads to the partial destruction of the equilibrium configuration in the inner accretion disk/corona region in the surrounds of the BH, suggests that this process could be related to the transition from  the 
low/hard to the high/soft steep-power-law (SPL) X-ray states often detected in microquasars  \citep{remillard_mcclintock_06,zhang_etal_14}, as argued in GPK10. However, a detailed analysis of this transition and the accretion disk-coronal state that follows an event of fast reconnection, as well as the reproduction of the whole spectral energy distribution of the sources is out of the scope of this work. 

Our results change considerably in the case of blazars. Although after Doppler beaming correction, in most cases their radio emission lies within the upper part of the magnetic reconnection power $\dot{W}_{B}$ diagram, which corresponds to accretion rates much larger than $\dot{m}=0.05$, this emission, in general, does not seem to follow the same trend as that of the LLAGNs or of the magnetic reconnection power itself (Figure \ref{fig:radio_radio}). In the case of their gamma-ray luminosity, even with the Doppler beaming correction, most of this power lies well above the $\dot{W}_{B}$ diagram (Figure \ref{fig:radio_gamma}). Comparing this with the gamma-ray emission of a sample of GRBs, we see that the line that links both the blazars and the GRBs gamma-ray emissions in the diagram is anti-correlated with the gray zone corresponding to the fast magnetic reconnection power in the core. 
This suggests that the emission in these sources cannot be attributed to this acceleration process. 
In the case of blazars, for which the jet axis points to the line of sight, this result is consistent with the standard scenario for these sources where the emission is attributed to relativistic particle acceleration along the jet which has relativistic bulk velocities.
A similar scenario is applicable to GRBs. Their prompt gamma-ray emission and the afterglow radio emission are attributed to the reverse internal shock and the forward external shock, respectively, of a super relativistically beamed jet after it breaks out from the stellar progenitor envelope \citep[e.g.,][]{woosley_93, paczynski_98, macFadyen_woosley_99}. 
In other words, in both classes of sources, the observed emission seems to be produced at the jet at distances larger than a few $20 R_S$ from the core of the sources and probably cannot be explained by the magnetic reconnection scenario here described. Any deep core emission in these sources is probably screened by opacity and by the jet pointing towards our line of sight.  
This result is consistent with the predictions of GPK10, and also with \citet{nemmen_etal_12} whose observed correlation between GRBs and blazars suggests that the gamma-ray and radio emission from such sources is originated further out at the associated relativistic jet. 

%-----------------------------------------------------------------------------------------------------

We note that in recent work \citet{zhang_yan_11} invoked the LV99 fast reconnection model and the GL05 first-order Fermi acceleration mechanism to explain the emission features in some GRBs. Specifically, they suggested that the GRB prompt emission would occur in a Poynting-flux dominated regime through the collision of multiple injected shells into the jet flow. These would distort the magnetic  field lines and induce fast reconnection which in turn would induce turbulence further distorting the magnetic field lines, easing additional magnetic reconnection and resulting in a runway release of the stored magnetic field energy and particle acceleration. This mechanism is somewhat similar to what we have suggested here in the sense that the onset of instabilities and the continuous reconnection during the building of the corona itself may trigger turbulence which in turn speeds up the overall process. However, distinctly from  \citet{zhang_yan_11} model which is a mechanism occurring in the jet beam, our model occurs within the nuclear region of the system. Other works have also investigated the effects of fast reconnection along the jet \citep[e.g.,][]{giannios_10, uzdensky_mcKinney_11, dgdp_etal_13}. In particular, \citet{dgdp_etal_13} have injected test particles in an MHD simulation of a  relativistic jet and found evidence of efficient particle acceleration by reconnection in this system.

We should stress that the results of the fast magnetic reconnection model investigated here are relatively insensitive to the physical parameters inside the accretion disk, except for the accretion rate which was allowed to vary between $0.0005 \leq \dot{m} \leq 1$. The source mass is the more critical parameter in our model as it varies over $10^{10}$ orders of magnitude. 
 This trend seems to be confirmed when we repeat the calculations here presented but adopting a magnetically dominated advective flow \citep[MDAF or MAD, see][]{meier_12, sikora_begelman_13} to describe the disk accretion/corona in the inner region of the source, rather than the standard Shakura-Sunyaev disk. In fact, with this new approach we find that the results and the slope of the diagrams of Figures \ref{fig:graf_radio} to \ref{fig:radio_gamma} do not change substantially \citep[see][]{singh_etal_14}. 
 
 As remarked, other contemporary works have also explored magnetic processes in the surrounds of BH systems and other compact sources to explain their emission \citep[see, e.g.,][]{soker_10, cemeljic_etal_13, uzdensky_spitkovsky_14, huang_etal_14}.
In particular \citet{huang_etal_14} investigated magnetic reconnection in the
surrounds of BH binary systems (microquasars) employing a radiatively inefficient advection-dominated accretion flow (RIAF) to describe the accretion disk combined with a jet model. 
They find that this  could explain the observed correlation between the radio and X-ray emission for the high soft state. Furthermore, they argue  that the sources that deviate from this correlation (the outliers) could be explained by an appropriate combination of these two mechanisms, a result that is consistent with the present analysis and the arguing of GPK10.  

\subsection{Summary and Final remarks}

In summary, the results of the present work indicate that  in the case of microquasars and low luminosity AGNs (LLAGNs), the power released by fast magnetic reconnection  driven by turbulence in the surrounds of the BH is able to explain  the observed  core radio and gamma-ray emission of these sources, therefore  indicating that the surrounds of the BHs (as sketched in Figure \ref{fig:scheme})  can be the acceleration region in these cases.  Also, according to our results, fast reconnection induced by anomalous resistivity is clearly less efficient to provide the appropriate magnetic power for most of the sources of the sample, therefore, fast reconnection induced by turbulence (as described in LV99) is clearly more appropriate and besides, it results nearly the same trend (slope) of the observed luminosity distributions for these sources (see Figures \ref{fig:radio01} to \ref{fig:radio_gamma00}).
On the other hand, in the case of blazars (and GRBs), our results show that the magnetic power released by fast reconnection (either driven by turbulence or anomalous resistivity) in the surrounds of the central source is clearly not sufficient to explain both the observed radio and gamma-ray radiation for most of these sources (Figures \ref{fig:radio_radio} and \ref{fig:radio_gamma}). This is probably due to the fact that these sources have their jets pointing to our line of sight and therefore, the core emission is screened by the jet. So that what is effectively observed is emission coming from further out - from the jet, as generally expected for these sources.

 The results above connecting both the radio and gamma-ray emission from low-luminosity compact sources to magnetically dominated reconnection processes are very promising as they suggest a unifying single process of relativistic particle acceleration in the core region which may naturally help with the interpretation of the observed correlations of LLAGNs and microquasars, as remarked, and also with clues for  existing unification AGN theories, providing important predictions for the coming new generation of VHE observatories with much larger sensitivity and energy range to perform emission and variability studies, such as the Cherenkov Telescope Array \citep[CTA; see][]{actis_etal_11, acharya_etal_13, sol_etal_13}. Also, multi-frequency observation \citep[as, e.g.,][]{hovatta_etal_14} will be crucial to better constrain the location of the gamma-ray emission and the acceleration mechanisms.  
 
 Finally, we should note that in this work we have focussed on the total power released by magnetic reconnection in the core region of the sources, without examining the radiation mechanisms by which this energy can be transformed into radio or gamma-ray emission. In a companion work we have explored the acceleration mechanism above operating in the core region of the microquasars Cyg X-$1$ and Cyg X-$3$ (see Figures \ref{fig:graf_radio}, \ref{fig:radio_gamma00} and \ref{fig:radio_gamma}) and have reproduced their entire observed non-thermal spectral energy distribution (SED), from the radio to the gamma-ray flux profile \citep[see][]{khiali_etal_14}. 

\acknowledgments
 This work has been partially supported by grants from the Brazilian agencies FAPESP (2013/09065-8, 2013/10559-5), CNPq (142220/2013-2 and 306598/2009-4) and CAPES. This paper has also benefited from very fruitful discussions with Alex Lazarian, Zulema Abraham, Rodrigo Nemmen, Tsvi Piran, and James Stone. The authors also acknowledge the useful comments from an anonymous referee.

\begin{appendix}
\section{Gas and radiation pressure regimes}
\label{app:prad_pgas}

\begin{figure}[t]
 \centering
 \includegraphics[width=0.5\textwidth] {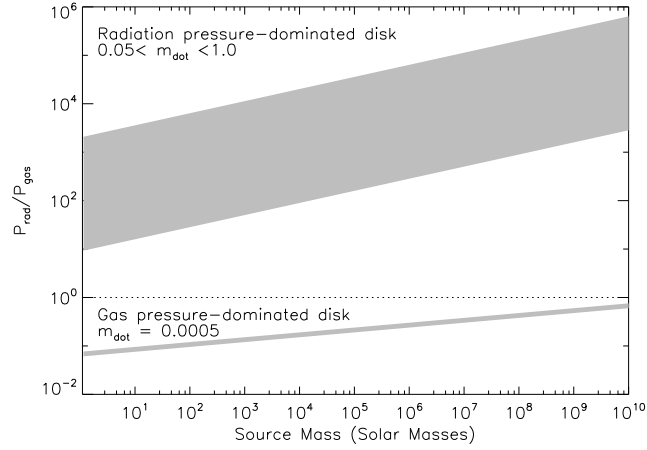}
 \caption{Ratio between the radiation and the gas pressure of the accretion disk ($P_{rad}/P_{gas}$) as a function of the source mass for different values of $\dot{m}$. The gray regions give the parametric space for $P_{rad}/P_{gas}$ for the radiation pressure (top) and the gas pressure dominated (bottom) regimes. The parametric space considered is $0.05 \leq \alpha \leq 0.5$, $\dot{m} \simeq 5\times10^{-4}$ for the gas-pressure dominated regime, and $5\times10^{-2} < \dot{m} < 1$ for the radiation-pressure dominated regime.}
 \label{fig:pp}
\end{figure}

Figure \ref{fig:pp} presents the ratio between the radiation and the gas pressure in the accretion disk versus the source mass for different values of the accretion rate, as obtained from equations (\ref{eq:pgas}) and (\ref{eq:prad}). 

\section{Supplementary tables}
\label{app:tables}

\begin{longtable*}{lcccccccccc}
\caption{Radio and gamma-ray emission (when available) of LLAGNs and microquasars (or galactic black hole binaries).}\\ 
\hline \hline \\[-2ex]
\multicolumn{1}{c}{Sources} &
\multicolumn{1}{c}{Type} &
\multicolumn{1}{c}{$\log_{10}(L_{R}[erg/s])$} &
\multicolumn{1}{c}{Notes} &
\multicolumn{1}{c}{Ref.} &
\multicolumn{1}{c}{$\log_{10}(L_{\gamma}[erg/s])$} &
\multicolumn{1}{c}{Notes} &
\multicolumn{1}{c}{Ref.} &
\multicolumn{1}{c}{$\log_{10}(M/M_{\odot})$} &
\multicolumn{1}{c}{Ref.}
\\
\multicolumn{1}{c}{(1)} &
\multicolumn{1}{c}{(2)} &
\multicolumn{1}{c}{(3)} &
\multicolumn{1}{c}{(4)} &
\multicolumn{1}{c}{(5)} &
\multicolumn{1}{c}{(6)} &
\multicolumn{1}{c}{(7)} &
\multicolumn{1}{c}{(8)} &
\multicolumn{1}{c}{(9)} &
\multicolumn{1}{c}{(10)}

\\[0.5ex] \hline
\\[-1.8ex]

\endfirsthead

\hline \hline\\[-2ex]

\multicolumn{1}{c}{Sources} &
\multicolumn{1}{c}{Type} &
\multicolumn{1}{c}{$\log_{10}(L_{R}[erg/s])$} &
\multicolumn{1}{c}{Notes} &
\multicolumn{1}{c}{Ref.} &
\multicolumn{1}{c}{$\log_{10}(L_{\gamma}[erg/s])$} &
\multicolumn{1}{c}{Notes} &
\multicolumn{1}{c}{Ref.} &
\multicolumn{1}{c}{$\log_{10}(M/M_{\odot})$} &
\multicolumn{1}{c}{Ref.}
\\
\multicolumn{1}{c}{(1)} &
\multicolumn{1}{c}{(2)} &
\multicolumn{1}{c}{(3)} &
\multicolumn{1}{c}{(4)} &
\multicolumn{1}{c}{(5)} &
\multicolumn{1}{c}{(6)} &
\multicolumn{1}{c}{(7)} &
\multicolumn{1}{c}{(8)} &
\multicolumn{1}{c}{(9)} &
\multicolumn{1}{c}{(10)}

\\[0.5ex] \hline
\\[-1.8ex]

\endhead

\multicolumn{7}{l}{{\footnotesize{Continued on next page}}}\\
\endfoot
\hline

\endlastfoot

IC 239	&	L	&	$	36.66	$	&	UL	&	[0]	&	$		$	&		&		&	$	6.67	$	&	[3][4]	\\
IC 356	&	T	&	$	36.77	$	&	UL	&	[0]	&	$		$	&		&		&	$	7.69	$	&	[3][4]	\\
IC 520	&	T	&	$	37.60	$	&	UL	&	[0]	&	$		$	&		&		&	$	7.44	$	&	[3][4]	\\
IC 1727	&	T/L	&	$	36.04	$	&	UL	&	[0]	&	$		$	&		&		&	$	7.41	$	&	[3][4]	\\
NGC 185	&	S	&	$	33.99	$	&	UL	&	[0]	&	$		$	&		&		&	$	3.89	$	&	[3][4]	\\
NGC 266	&	L	&	$	38.46	$	&		&	[0]	&	$		$	&		&		&	$	8.39	$	&	[3][4]	\\
NGC 315	&	L	&	$	40.57	$	&		&	[0]	&	$		$	&		&		&	$	8.96	$	&	[3][4][5]	\\
NGC 404	&	L	&	$	35.13	$	&	UL	&	[0]	&	$		$	&		&		&	$	4.84	$	&	[3][4]	\\
NGC 410	&	L	&	$	37.96	$	&	UL	&	[0]	&	$		$	&		&		&	$	8.88	$	&	[3][4]	\\
NGC 428	&	T/L	&	$	36.56	$	&	UL	&	[0]	&	$		$	&		&		&	$	4.36	$	&	[3][4]	\\
NGC 474	&	L	&	$	37.46	$	&	UL	&	[0]	&	$		$	&		&		&	$	7.66	$	&	[3][4]	\\
NGC 488	&	T	&	$	37.19	$	&	UL	&	[0]	&	$		$	&		&		&	$	8.06	$	&	[3][4]	\\
NGC 521	&	T/H	&	$	37.91	$	&	UL	&	[0]	&	$		$	&		&		&	$	8.23	$	&	[3][4]	\\
NGC 524	&	T	&	$	37.45	$	&		&	[0]	&	$		$	&		&		&	$	8.48	$	&	[3][4]	\\
NGC 660	&	T/H	&	$	36.36	$	&	UL	&	[0]	&	$		$	&		&		&	$	7.10	$	&	[3][4]	\\
NGC 676	&	S	&	$	37.01	$	&	UL	&	[0]	&	$		$	&		&		&	$	7.55	$	&	[3][4]	\\
NGC 718	&	L	&	$	37.09	$	&	UL	&	[0]	&	$		$	&		&		&	$	7.15	$	&	[3][4]	\\
NGC 777	&	S/L	&	$	38.08	$	&	UL	&	[0]	&	$		$	&		&		&	$	9.01	$	&	[3][4]	\\
NGC 841	&	L	&	$	37.98	$	&	UL	&	[0]	&	$		$	&		&		&	$	7.70	$	&	[3][4]	\\
NGC 1055	&	T/L	&	$	36.71	$	&	UL	&	[0]	&	$		$	&		&		&	$	6.41	$	&	[3][4]	\\
NGC 1058	&	S	&	$	36.13	$	&	UL	&	[0]	&	$		$	&		&		&	$	5.47	$	&	[3][4]	\\
NGC 1167	&	S	&	$	39.54	$	&		&	[0]	&	$		$	&		&		&	$	8.16	$	&	[3][4]	\\
NGC 1169	&	L	&	$	37.49	$	&	UL	&	[0]	&	$		$	&		&		&	$	7.93	$	&	[3][4]	\\
NGC 1275 (Per A)	&	S	&	$	41.42	$	&		&	[0]	&	$	41.63	$	&		&	[13]	&	$	8.62	$	&	[3][4][5]	\\
NGC 1275 (Per A)	&	S	&	$	41.42	$	&		&	[0]	&	$	43.22	$	&		&	[13]	&	$	8.62	$	&	[3][4][5]	\\
NGC 1961	&	L	&	$	37.88	$	&	UL	&	[0]	&	$		$	&		&		&	$	8.50	$	&	[3][4]	\\
NGC 2273	&	S	&	$	37.78	$	&		&	[0]	&	$		$	&		&		&	$	7.43	$	&	[3][4][5]	\\
NGC 2336	&	L/S	&	$	37.49	$	&	UL	&	[0]	&	$		$	&		&		&	$	7.24	$	&	[3][4]	\\
NGC 2541	&	T/H	&	$	36.31	$	&	UL	&	[0]	&	$		$	&		&		&	$	5.65	$	&	[3][4]	\\
NGC 2655	&	S	&	$	37.81	$	&		&	[0]	&	$		$	&		&		&	$	7.71	$	&	[3][4]	\\
NGC 2681	&	L	&	$	36.66	$	&	UL	&	[0]	&	$		$	&		&		&	$	7.21	$	&	[3][4]	\\
NGC 2683	&	L/S	&	$	35.72	$	&	UL	&	[0]	&	$		$	&		&		&	$	7.21	$	&	[3][4]	\\
NGC 2685	&	S/T	&	$	36.63	$	&	UL	&	[0]	&	$		$	&		&		&	$	6.82	$	&	[3][4]	\\
NGC 2768	&	L	&	$	37.91	$	&		&	[0]	&	$		$	&		&		&	$	7.98	$	&	[3][4]	\\
NGC 2787	&	L	&	$	37.33	$	&		&	[0]	&	$		$	&		&		&	$	7.97	$	&	[3][4][5]	\\
NGC 2832	&	L	&	$	38.36	$	&	UL	&	[0]	&	$		$	&		&		&	$	9.09	$	&	[3][4]	\\
NGC 2841	&	L	&	$	36.46	$	&		&	[0]	&	$		$	&		&		&	$	8.31	$	&	[3][4][5]	\\
NGC 2859	&	T	&	$	37.07	$	&	UL	&	[0]	&	$		$	&		&		&	$	7.92	$	&	[3][4]	\\
NGC 2911	&	L	&	$	38.75	$	&		&	[0]	&	$		$	&		&		&	$	8.47	$	&	[3][4]	\\
NGC 2985	&	T	&	$	36.96	$	&	UL	&	[0]	&	$		$	&		&		&	$	7.84	$	&	[3][4]	\\
NGC 3031	&	S	&	$	37.59	$	&		&	[0]	&	$		$	&		&		&	$	7.73	$	&	[3][4][5]	\\
NGC 3079	&	S	&	$	38.56	$	&		&	[0]	&	$		$	&		&		&	$	7.83	$	&	[3][4][5]	\\
NGC 3147	&	S	&	$	38.38	$	&		&	[0]	&	$		$	&		&		&	$	8.55	$	&	[3][4][5]	\\
NGC 3166	&	L	&	$	37.12	$	&	UL	&	[0]	&	$		$	&		&		&	$	7.60	$	&	[3][4]	\\
NGC 3169	&	L	&	$	37.68	$	&		&	[0]	&	$		$	&		&		&	$	7.96	$	&	[3][4][5]	\\
NGC 3190	&	L	&	$	37.00	$	&		&	[0]	&	$		$	&		&		&	$	8.01	$	&	[3][4]	\\
NGC 3193	&	L	&	$	37.17	$	&	UL	&	[0]	&	$		$	&		&		&	$	8.08	$	&	[3][4]	\\
NGC 3226	&	L	&	$	37.70	$	&		&	[0]	&	$		$	&		&		&	$	8.14	$	&	[3][4][5]	\\
NGC 3227	&	S	&	$	37.43	$	&		&	[0]	&	$	41.30	$	&	UL	&	[8]	&	$	7.46	$	&	[3][4][5]	\\
NGC 3245	&	T	&	$	36.95	$	&	UL	&	[0]	&	$		$	&		&		&	$	8.21	$	&	[3][4]	\\
NGC 3254	&	S	&	$	37.18	$	&	UL	&	[0]	&	$		$	&		&		&	$	7.34	$	&	[3][4]	\\
NGC 3301	&	L	&	$	37.17	$	&	UL	&	[0]	&	$		$	&		&		&	$	7.21	$	&	[3][4]	\\
NGC 3368	&	L	&	$	36.07	$	&	UL	&	[0]	&	$		$	&		&		&	$	7.28	$	&	[3][4]	\\
NGC 3379	&	L/T	&	$	36.07	$	&	UL	&	[0]	&	$		$	&		&		&	$	8.06	$	&	[1][2][3][4]	\\
NGC 3414	&	L	&	$	37.41	$	&		&	[0]	&	$		$	&		&		&	$	8.46	$	&	[3][4]	\\
NGC 3433	&	L/T	&	$	37.63	$	&	UL	&	[0]	&	$		$	&		&		&	$	6.41	$	&	[3][4]	\\
NGC 3486	&	S	&	$	36.00	$	&	UL	&	[0]	&	$		$	&		&		&	$	5.85	$	&	[3][4]	\\
NGC 3489	&	T/S	&	$	35.87	$	&	UL	&	[0]	&	$		$	&		&		&	$	6.91	$	&	[3][4]	\\
NGC 3507	&	L	&	$	37.03	$	&	UL	&	[0]	&	$		$	&		&		&	$	6.54	$	&	[3][4]	\\
NGC 3516	&	S	&	$	37.55	$	&		&	[0]	&	$	41.70	$	&	UL	&	[8]	&	$	7.58	$	&	[3][4][5]	\\
NGC 3607	&	L	&	$	37.06	$	&		&	[0]	&	$		$	&		&		&	$	8.34	$	&	[3][4]	\\
NGC 3608	&	L/S	&	$	37.17	$	&	UL	&	[0]	&	$		$	&		&		&	$	8.08	$	&	[2][3][4]	\\
NGC 3623	&	L	&	$	35.94	$	&	UL	&	[0]	&	$		$	&		&		&	$	7.54	$	&	[3][4]	\\
NGC 3626	&	L	&	$	37.27	$	&	UL	&	[0]	&	$		$	&		&		&	$	7.48	$	&	[3][4]	\\
NGC 3627	&	T/S	&	$	35.94	$	&		&	[0]	&	$		$	&		&		&	$	7.26	$	&	[3][4][5]	\\
NGC 3628	&	T	&	$	36.21	$	&		&	[0]	&	$		$	&		&		&	$	6.45	$	&	[3][4]	\\
NGC 3642	&	L	&	$	37.24	$	&	UL	&	[0]	&	$		$	&		&		&	$	6.95	$	&	[3][4]	\\
NGC 3675	&	T	&	$	36.47	$	&	UL	&	[0]	&	$		$	&		&		&	$	7.00	$	&	[3][4][5]	\\
NGC 3681	&	T	&	$	37.03	$	&	UL	&	[0]	&	$		$	&		&		&	$	6.67	$	&	[3][4]	\\
NGC 3692	&	T	&	$	37.27	$	&	UL	&	[0]	&	$		$	&		&		&	$	7.07	$	&	[3][4]	\\
NGC 3705	&	T	&	$	36.79	$	&	UL	&	[0]	&	$		$	&		&		&	$	7.05	$	&	[3][4]	\\
NGC 3718	&	L	&	$	37.74	$	&		&	[0]	&	$		$	&		&		&	$	7.71	$	&	[3][4]	\\
NGC 3735	&	S	&	$	37.66	$	&	UL	&	[0]	&	$		$	&		&		&	$	7.46	$	&	[3][4]	\\
NGC 3780	&	L	&	$	37.44	$	&		&	[0]	&	$		$	&		&		&	$	6.63	$	&	[3][4]	\\
NGC 3898	&	T	&	$	36.94	$	&	UL	&	[0]	&	$		$	&		&		&	$	8.14	$	&	[3][4]	\\
NGC 3900	&	L	&	$	37.37	$	&	UL	&	[0]	&	$		$	&		&		&	$	7.45	$	&	[3][4]	\\
NGC 3917	&	T	&	$	36.81	$	&	UL	&	[0]	&	$		$	&		&		&	$	5.04	$	&	[3][4]	\\
NGC 3941	&	S	&	$	36.85	$	&	UL	&	[0]	&	$		$	&		&		&	$	7.34	$	&	[3][4]	\\
NGC 3945	&	L	&	$	37.22	$	&		&	[0]	&	$		$	&		&		&	$	7.97	$	&	[3][4]	\\
NGC 3953	&	T	&	$	36.82	$	&	UL	&	[0]	&	$		$	&		&		&	$	7.30	$	&	[3][4]	\\
NGC 3976	&	S	&	$	37.59	$	&	UL	&	[0]	&	$		$	&		&		&	$	8.03	$	&	[3][4]	\\
NGC 3982	&	S	&	$	36.72	$	&	UL	&	[0]	&	$		$	&		&		&	$	6.16	$	&	[3][4]	\\
NGC 3992	&	T	&	$	36.82	$	&	UL	&	[0]	&	$		$	&		&		&	$	7.62	$	&	[3][4]	\\
NGC 3998	&	L	&	$	38.68	$	&		&	[0]	&	$		$	&		&		&	$	8.72	$	&	[3][4][5]	\\
NGC 4013	&	T	&	$	36.72	$	&	UL	&	[0]	&	$		$	&		&		&	$	6.54	$	&	[3][4]	\\
NGC 4036	&	L	&	$	37.22	$	&	UL	&	[0]	&	$		$	&		&		&	$	8.15	$	&	[3][4]	\\
NGC 4051	&	S	&	$	36.72	$	&	UL	&	[0]	&	$	40.60	$	&	UL	&	[8]	&	$	6.63	$	&	[3][4][5]	\\
NGC 4111	&	L	&	$	36.82	$	&	UL	&	[0]	&	$		$	&		&		&	$	7.57	$	&	[3][4]	\\
NGC 4125	&	T	&	$	37.03	$	&	UL	&	[0]	&	$		$	&		&		&	$	8.48	$	&	[3][4]	\\
NGC 4138	&	S	&	$	36.89	$	&		&	[0]	&	$	40.60	$	&	UL	&	[8]	&	$	7.28	$	&	[3][4]	\\
NGC 4143	&	L	&	$	37.24	$	&		&	[0]	&	$		$	&		&		&	$	8.21	$	&	[3][4][5]	\\
NGC 4150	&	T	&	$	36.23	$	&	UL	&	[0]	&	$		$	&		&		&	$	6.50	$	&	[3][4]	\\
NGC 4151	&	S	&	$	37.97	$	&		&	[0]	&	$	40.30	$	&	UL	&	[8]	&	$	7.00	$	&	[3][4][5]	\\
NGC 4168	&	S	&	$	37.19	$	&		&	[0]	&	$		$	&		&		&	$	7.96	$	&	[3][4]	\\
NGC 4169	&	S	&	$	37.74	$	&		&	[0]	&	$		$	&		&		&	$	7.97	$	&	[3][4]	\\
NGC 4192	&	T	&	$	36.82	$	&	UL	&	[0]	&	$		$	&		&		&	$	7.37	$	&	[3][4]	\\
NGC 4203	&	L	&	$	37.21	$	&		&	[0]	&	$		$	&		&		&	$	7.80	$	&	[3][4][5]	\\
NGC 4216	&	T	&	$	36.79	$	&		&	[0]	&	$		$	&		&		&	$	8.14	$	&	[3][4]	\\
NGC 4220	&	T	&	$	36.88	$	&	UL	&	[0]	&	$		$	&		&		&	$	7.02	$	&	[3][4]	\\
NGC 4258	&	S	&	$	36.34	$	&		&	[0]	&	$		$	&		&		&	$	7.53	$	&	[1][2][3][4][5]	\\
NGC 4261	&	L	&	$	39.83	$	&		&	[0]	&	$		$	&		&		&	$	8.80	$	&	[1][2][3][4][5]	\\
NGC 4278	&	L	&	$	38.18	$	&		&	[0]	&	$		$	&		&		&	$	8.86	$	&	[3][4][5]	\\
NGC 4281	&	T	&	$	37.35	$	&	UL	&	[0]	&	$		$	&		&		&	$	8.61	$	&	[3][4]	\\
NGC 4293	&	L	&	$	36.56	$	&		&	[0]	&	$		$	&		&		&	$	7.13	$	&	[3][4]	\\
NGC 4314	&	L	&	$	36.23	$	&	UL	&	[0]	&	$		$	&		&		&	$	6.98	$	&	[3][4]	\\
NGC 4321	&	T	&	$	36.66	$	&	UL	&	[0]	&	$		$	&		&		&	$	6.70	$	&	[3][4][5]	\\
NGC 4324	&	T	&	$	37.35	$	&	UL	&	[0]	&	$		$	&		&		&	$	6.75	$	&	[3][4]	\\
NGC 4346	&	L	&	$	36.72	$	&	UL	&	[0]	&	$		$	&		&		&	$	7.45	$	&	[3][4]	\\
NGC 4350	&	T	&	$	36.66	$	&	UL	&	[0]	&	$		$	&		&		&	$	7.96	$	&	[3][4]	\\
NGC 4374	&	L	&	$	38.97	$	&		&	[0]	&	$		$	&		&		&	$	9.07	$	&	[1][2][3][4][5]	\\
NGC 4378	&	S	&	$	37.52	$	&	UL	&	[0]	&	$		$	&		&		&	$	8.06	$	&	[3][4]	\\
NGC 4388	&	S	&	$	37.05	$	&		&	[0]	&	$	40.70	$	&	UL	&	[8]	&	$	6.80	$	&	[3][4][5]	\\
NGC 4394	&	L	&	$	36.66	$	&	UL	&	[0]	&	$		$	&		&		&	$	7.19	$	&	[3][4]	\\
NGC 4395	&	S	&	$	35.32	$	&	UL	&	[0]	&	$	39.60	$	&	UL	&	[8]	&	$	4.83	$	&	[1][3][4][5]	\\
NGC 4414	&	T	&	$	36.19	$	&	UL	&	[0]	&	$		$	&		&		&	$	7.02	$	&	[3][4]	\\
NGC 4419	&	T	&	$	37.14	$	&		&	[0]	&	$		$	&		&		&	$	6.96	$	&	[3][4]	\\
NGC 4429	&	T	&	$	36.75	$	&	UL	&	[0]	&	$		$	&		&		&	$	7.90	$	&	[3][4]	\\
NGC 4435	&	T/H	&	$	36.75	$	&	UL	&	[0]	&	$		$	&		&		&	$	7.68	$	&	[3][4]	\\
NGC 4438	&	L	&	$	36.66	$	&	UL	&	[0]	&	$		$	&		&		&	$	7.40	$	&	[3][4]	\\
NGC 4450	&	L	&	$	37.01	$	&		&	[0]	&	$		$	&		&		&	$	7.35	$	&	[3][4][5]	\\
NGC 4457	&	L	&	$	36.74	$	&	UL	&	[0]	&	$		$	&		&		&	$	7.02	$	&	[3][4][5]	\\
NGC 4459	&	T	&	$	36.71	$	&	UL	&	[0]	&	$		$	&		&		&	$	7.86	$	&	[3][4]	\\
NGC 4472	&	S	&	$	37.28	$	&		&	[0]	&	$		$	&		&		&	$	8.78	$	&	[3][4][5]	\\
NGC 4477	&	S	&	$	36.71	$	&	UL	&	[0]	&	$		$	&		&		&	$	7.87	$	&	[3][4]	\\
NGC 4486 (M87)	&	L	&	$	40.14	$	&		&	[0]	&	$	39.91	$	&		&	[10]	&	$	9.33	$	&	[1][3][4][5]	\\
NGC 4486 (M87)	&	L	&	$	40.14	$	&		&	[0]	&	$	41.09	$	&		&	[10]	&	$	9.33	$	&	[1][3][4][5]	\\
NGC 4494	&	L	&	$	36.13	$	&	UL	&	[0]	&	$		$	&		&		&	$	7.60	$	&	[3][4][5]	\\
NGC 4501	&	S	&	$	36.75	$	&	UL	&	[0]	&	$		$	&		&		&	$	7.83	$	&	[3][4][5]	\\
NGC 4527	&	T	&	$	36.56	$	&	UL	&	[0]	&	$		$	&		&		&	$	7.38	$	&	[3][4]	\\
NGC 4548	&	L	&	$	36.79	$	&		&	[0]	&	$		$	&		&		&	$	7.32	$	&	[3][4][5]	\\
NGC 4550	&	L	&	$	36.55	$	&		&	[0]	&	$		$	&		&		&	$	6.88	$	&	[3][4]	\\
NGC 4552	&	T	&	$	38.47	$	&		&	[0]	&	$		$	&		&		&	$	8.57	$	&	[3][4]	\\
NGC 4565	&	S	&	$	36.80	$	&		&	[0]	&	$		$	&		&		&	$	7.64	$	&	[3][4][5]	\\
NGC 4569	&	T	&	$	36.75	$	&	UL	&	[0]	&	$		$	&		&		&	$	7.45	$	&	[3][4]	\\
NGC 4579	&	S/L	&	$	38.15	$	&		&	[0]	&	$		$	&		&		&	$	7.81	$	&	[3][4][5]	\\
NGC 4589	&	L	&	$	38.28	$	&		&	[0]	&	$		$	&		&		&	$	8.35	$	&	[3][4]	\\
NGC 4596	&	L	&	$	36.75	$	&	UL	&	[0]	&	$		$	&		&		&	$	7.54	$	&	[3][4]	\\
NGC 4636	&	L	&	$	36.92	$	&		&	[0]	&	$		$	&		&		&	$	8.09	$	&	[3][4][5]	\\
NGC 4639	&	S	&	$	36.75	$	&	UL	&	[0]	&	$		$	&		&		&	$	6.60	$	&	[3][4]	\\
NGC 4643	&	T	&	$	37.08	$	&	UL	&	[0]	&	$		$	&		&		&	$	7.58	$	&	[3][4]	\\
NGC 4651	&	L	&	$	36.75	$	&	UL	&	[0]	&	$		$	&		&		&	$	6.84	$	&	[3][4]	\\
NGC 4698	&	S	&	$	36.71	$	&	UL	&	[0]	&	$		$	&		&		&	$	7.48	$	&	[3][4]	\\
NGC 4713	&	T	&	$	36.81	$	&	UL	&	[0]	&	$		$	&		&		&	$	4.14	$	&	[3][4]	\\
NGC 4725	&	S	&	$	36.40	$	&	UL	&	[0]	&	$		$	&		&		&	$	7.40	$	&	[3][4][5]	\\
NGC 4736	&	L	&	$	35.80	$	&		&	[0]	&	$		$	&		&		&	$	7.12	$	&	[3][4][5]	\\
NGC 4750	&	L	&	$	37.27	$	&	UL	&	[0]	&	$		$	&		&		&	$	7.40	$	&	[3][4]	\\
NGC 4762	&	L	&	$	36.66	$	&		&	[0]	&	$		$	&		&		&	$	7.54	$	&	[3][4]	\\
NGC 4772	&	L	&	$	37.20	$	&		&	[0]	&	$		$	&		&		&	$	7.55	$	&	[3][4]	\\
NGC 4826	&	T	&	$	35.44	$	&	UL	&	[0]	&	$		$	&		&		&	$	6.84	$	&	[3][4]	\\
NGC 4866	&	L	&	$	36.71	$	&	UL	&	[0]	&	$		$	&		&		&	$	8.20	$	&	[3][4]	\\
NGC 5005	&	L	&	$	36.96	$	&	UL	&	[0]	&	$		$	&		&		&	$	7.84	$	&	[3][4]	\\
NGC 5012	&	T	&	$	37.47	$	&	UL	&	[0]	&	$		$	&		&		&	$	7.46	$	&	[3][4]	\\
NGC 5033	&	S	&	$	36.95	$	&		&	[0]	&	$		$	&		&		&	$	7.36	$	&	[3][4][5]	\\
NGC 5055	&	T	&	$	36.01	$	&	UL	&	[0]	&	$		$	&		&		&	$	6.86	$	&	[3][4]	\\
NGC 5194	&	S	&	$	36.07	$	&	UL	&	[0]	&	$		$	&		&		&	$	6.74	$	&	[3][4][5]	\\
NGC 5195	&	L	&	$	36.24	$	&	UL	&	[0]	&	$		$	&		&		&	$	7.28	$	&	[3][4]	\\
NGC 5273	&	S	&	$	36.96	$	&	UL	&	[0]	&	$		$	&		&		&	$	6.20	$	&	[3][4][5]	\\
NGC 5297	&	L	&	$	37.59	$	&	UL	&	[0]	&	$		$	&		&		&	$	6.60	$	&	[3][4]	\\
NGC 5322	&	L	&	$	38.36	$	&		&	[0]	&	$		$	&		&		&	$	8.43	$	&	[3][4]	\\
NGC 5353	&	L/T	&	$	38.68	$	&		&	[0]	&	$		$	&		&		&	$	8.82	$	&	[3][4]	\\
NGC 5354	&	T/L	&	$	38.28	$	&		&	[0]	&	$		$	&		&		&	$	8.27	$	&	[3][4]	\\
NGC 5363	&	L	&	$	38.54	$	&		&	[0]	&	$		$	&		&		&	$	8.33	$	&	[3][4]	\\
NGC 5371	&	L	&	$	37.59	$	&	UL	&	[0]	&	$		$	&		&		&	$	7.92	$	&	[3][4]	\\
NGC 5377	&	L	&	$	37.72	$	&		&	[0]	&	$		$	&		&		&	$	7.87	$	&	[3][4]	\\
NGC 5395	&	S/L	&	$	37.77	$	&	UL	&	[0]	&	$		$	&		&		&	$	7.53	$	&	[3][4]	\\
NGC 5448	&	L	&	$	37.46	$	&	UL	&	[0]	&	$		$	&		&		&	$	7.24	$	&	[3][4]	\\
NGC 5485	&	L	&	$	37.47	$	&	UL	&	[0]	&	$		$	&		&		&	$	8.08	$	&	[3][4]	\\
NGC 5566	&	L	&	$	37.14	$	&	UL	&	[0]	&	$		$	&		&		&	$	7.70	$	&	[3][4]	\\
NGC 5631	&	S/L	&	$	37.46	$	&	UL	&	[0]	&	$		$	&		&		&	$	7.77	$	&	[3][4]	\\
NGC 5656	&	T	&	$	37.52	$	&	UL	&	[0]	&	$		$	&		&		&	$	7.11	$	&	[3][4]	\\
NGC 5678	&	T	&	$	37.36	$	&	UL	&	[0]	&	$		$	&		&		&	$	7.35	$	&	[3][4]	\\
NGC 5701	&	T	&	$	37.13	$	&	UL	&	[0]	&	$		$	&		&		&	$	7.22	$	&	[3][4]	\\
NGC 5746	&	T	&	$	37.19	$	&	UL	&	[0]	&	$		$	&		&		&	$	8.02	$	&	[3][4]	\\
NGC 5813	&	L	&	$	37.51	$	&		&	[0]	&	$		$	&		&		&	$	8.45	$	&	[3][4]	\\
NGC 5838	&	T	&	$	37.37	$	&		&	[0]	&	$		$	&		&		&	$	8.74	$	&	[3][4]	\\
NGC 5846	&	T	&	$	37.97	$	&		&	[0]	&	$		$	&		&		&	$	8.44	$	&	[3][4]	\\
NGC 5866	&	T	&	$	37.48	$	&		&	[0]	&	$		$	&		&		&	$	7.78	$	&	[3][4]	\\
NGC 5879	&	T/L	&	$	36.75	$	&	UL	&	[0]	&	$		$	&		&		&	$	6.45	$	&	[3][4]	\\
NGC 5921	&	T	&	$	37.06	$	&	UL	&	[0]	&	$		$	&		&		&	$	6.50	$	&	[3][4]	\\
NGC 6340	&	L	&	$	37.12	$	&	UL	&	[0]	&	$		$	&		&		&	$	7.48	$	&	[3][4]	\\
NGC 6384	&	T	&	$	37.11	$	&	UL	&	[0]	&	$		$	&		&		&	$	7.25	$	&	[3][4]	\\
NGC 6482	&	T/S	&	$	37.70	$	&	UL	&	[0]	&	$		$	&		&		&	$	9.04	$	&	[3][4]	\\
NGC 6500	&	L	&	$	39.38	$	&		&	[0]	&	$		$	&		&		&	$	8.23	$	&	[3][4][5]	\\
NGC 6503	&	T/S	&	$	35.83	$	&	UL	&	[0]	&	$		$	&		&		&	$	5.39	$	&	[3][4]	\\
NGC 6703	&	L	&	$	37.54	$	&	UL	&	[0]	&	$		$	&		&		&	$	7.90	$	&	[3][4]	\\
NGC 6951	&	S	&	$	37.20	$	&	UL	&	[0]	&	$		$	&		&		&	$	7.11	$	&	[3][4]	\\
NGC 7177	&	T	&	$	36.82	$	&	UL	&	[0]	&	$		$	&		&		&	$	7.28	$	&	[3][4]	\\
NGC 7217	&	L	&	$	36.41	$	&	UL	&	[0]	&	$		$	&		&		&	$	7.41	$	&	[3][4]	\\
NGC 7331	&	T	&	$	36.61	$	&	UL	&	[0]	&	$		$	&		&		&	$	7.41	$	&	[3][4]	\\
NGC 7479	&	S	&	$	37.66	$	&		&	[0]	&	$		$	&		&		&	$	7.60	$	&	[3][4]	\\
NGC 7626	&	L	&	$	39.18	$	&		&	[0]	&	$		$	&		&		&	$	8.68	$	&	[3][4]	\\
NGC 7742	&	T/S	&	$	36.99	$	&	UL	&	[0]	&	$		$	&		&		&	$	6.45	$	&	[3][4]	\\
NGC 7743	&	S	&	$	37.03	$	&		&	[0]	&	$		$	&		&		&	$	6.50	$	&	[3][4][5]	\\
NGC 7814	&	L	&	$	36.66	$	&	UL	&	[0]	&	$		$	&		&		&	$	7.83	$	&	[3][4]	\\
Ark 564	&	S	&	$	38.59	$	&		&	[5]	&	$		$	&		&		&	$	6.50	$	&	[5]	\\
Cyg A	&	S/L	&	$	41.43	$	&		&	[5]	&	$		$	&		&		&	$	9.40	$	&	[5]	\\
Fairall 9	&	S	&	$	37.68	$	&	UL	&	[5]	&	$	42.70	$	&	UL	&	[8]	&	$	7.91	$	&	[5]	\\
IC 1459	&	L	&	$	39.71	$	&		&	[5]	&	$		$	&		&		&	$	8.88	$	&	[2][3][4][5]	\\
IC 4296	&	L	&	$	39.68	$	&		&	[5]	&	$		$	&		&		&	$	9.10	$	&	[3][4][5]	\\
IC 4329A	&	S	&	$	38.94	$	&		&	[5]	&	$	42.40	$	&	UL	&	[8]	&	$	8.26	$	&	[3][4][5]	\\
Mrk 3	&	S	&	$	39.86	$	&		&	[5]	&	$		$	&		&		&	$	8.81	$	&	[5]	\\
Mrk 78	&	S	&	$	39.86	$	&		&	[5]	&	$		$	&		&		&	$	7.92	$	&	[5]	\\
Mrk 279	&	S	&	$	38.78	$	&		&	[5]	&	$	42.70	$	&	UL	&	[8]	&	$	7.62	$	&	[5]	\\
Mrk 335	&	S	&	$	38.27	$	&		&	[5]	&	$		$	&		&		&	$	6.79	$	&	[5]	\\
Mrk 348	&	S	&	$	39.70	$	&		&	[5]	&	$	41.60	$	&	UL	&	[8]	&	$	7.17	$	&	[5]	\\
Mrk 478	&	S	&	$	38.75	$	&		&	[5]	&	$		$	&		&		&	$	7.30	$	&	[5]	\\
Mrk 507	&	S	&	$	38.78	$	&		&	[5]	&	$		$	&		&		&	$	7.10	$	&	[5]	\\
Mrk 509	&	S	&	$	38.30	$	&		&	[5]	&	$	42.60	$	&	UL	&	[8]	&	$	7.86	$	&	[5]	\\
Mrk 573	&	S	&	$	38.22	$	&		&	[5]	&	$		$	&		&		&	$	7.25	$	&	[5]	\\
Mrk 590	&	S	&	$	38.70	$	&		&	[5]	&	$		$	&		&		&	$	7.23	$	&	[5]	\\
Mrk 766	&	S	&	$	38.32	$	&		&	[5]	&	$		$	&		&		&	$	6.64	$	&	[5]	\\
Mrk 1066	&	S	&	$	38.68	$	&		&	[5]	&	$		$	&		&		&	$	6.93	$	&	[5]	\\
NGC 1052	&	L	&	$	39.86	$	&		&	[5]	&	$		$	&		&		&	$	8.22	$	&	[3][4][5]	\\
NGC 1068	&	S	&	$	39.12	$	&		&	[5]	&	$		$	&		&		&	$	7.78	$	&	[1][2][3][4][5]	\\
NGC 1365	&	S	&	$	38.80	$	&		&	[5]	&	$	40.90	$	&	UL	&	[8]	&	$	7.62	$	&	[3][4][5]	\\
NGC 1386	&	S	&	$	36.70	$	&		&	[5]	&	$		$	&		&		&	$	7.65	$	&	[3][4][5]	\\
NGC 1667	&	S	&	$	37.34	$	&		&	[5]	&	$		$	&		&		&	$	7.97	$	&	[3][4][5]	\\
NGC 2110	&	S	&	$	38.99	$	&		&	[5]	&	$	41.40	$	&	UL	&	[8]	&	$	8.53	$	&	[3][4][5]	\\
NGC 2992	&	S	&	$	38.64	$	&		&	[5]	&	$		$	&		&		&	$	7.76	$	&	[3][4][5]	\\
NGC 3362	&	S	&	$	38.47	$	&		&	[5]	&	$		$	&		&		&	$	7.01	$	&	[3][4][5]	\\
NGC 4117	&	S	&	$	35.70	$	&		&	[5]	&	$		$	&		&		&	$	6.65	$	&	[3][4][5]	\\
NGC 4594	&	L	&	$	37.84	$	&		&	[5]	&	$		$	&		&		&	$	8.83	$	&	[1][3][4][5]	\\
NGC 5252	&	S	&	$	38.96	$	&		&	[5]	&	$	42.20	$	&	UL	&	[8]	&	$	8.10	$	&	[3][4][5]	\\
NGC 5347	&	S	&	$	37.10	$	&		&	[5]	&	$		$	&		&		&	$	6.70	$	&	[3][4][5]	\\
NGC 5548	&	S	&	$	38.58	$	&		&	[5]	&	$	42.10	$	&	UL	&	[8]	&	$	8.08	$	&	[3][4][5]	\\
NGC 5929	&	S	&	$	38.30	$	&		&	[5]	&	$		$	&		&		&	$	7.19	$	&	[3][4][5]	\\
NGC 6166	&	S	&	$	39.95	$	&		&	[5]	&	$		$	&		&		&	$	9.04	$	&	[3][4][5]	\\
NGC 6251	&	S	&	$	40.93	$	&		&	[5]	&	$		$	&		&		&	$	8.93	$	&	[2][3][4][5]	\\
NGC 7469	&	S	&	$	38.38	$	&		&	[5]	&	$	41.80	$	&	UL	&	[8]	&	$	7.24	$	&	[3][4][5]	\\
NGC 7672	&	S	&	$	37.25	$	&		&	[5]	&	$		$	&		&		&	$	6.80	$	&	[3][4][5]	\\
NGC 7682	&	S	&	$	38.88	$	&		&	[5]	&	$		$	&		&		&	$	7.25	$	&	[3][4][5]	\\
3C 120	&	S	&	$	41.55	$	&		&	[5]	&	$		$	&		&		&	$	7.36	$	&	[5]	\\
3C 390.3	&	S	&	$	41.09	$	&		&	[5]	&	$		$	&		&		&	$	8.53	$	&	[5]	\\
UGC 6100	&	S	&	$	38.50	$	&		&	[5]	&	$		$	&		&		&	$	7.72	$	&	[5]	\\
NGC 5128 (Cen A)	&	S	&	$	38.67	$	&		&	[11]	&	$	38.49	$	&		&	[12]	&	$	7.70	$	&	[11]	\\
NGC 5128 (Cen A)	&	S	&	$	38.67	$	&		&	[11]	&	$	40.61	$	&		&	[12]	&	$	7.70	$	&	[11]	\\
IC 310	&	Galaxy	&	$	39.77	$	&		&	[15]	&	$	41.70	$	&		&	[14]	&	$	8.30	$	&	[14]	\\
IC 310	&	Galaxy	&	$	39.77	$	&		&	[15]	&	$	43.30	$	&		&	[14]	&	$	8.30	$	&	[14]	\\
Cgy-X1	&	GBH	&	$	29.66	$	&		&	[5]	&	$	32.45	$	&		&	[16]	&	$	1.06	$	&	[5]	\\
Cgy-X1	&	GBH	&	$	29.66	$	&		&	[5]	&	$	33.56	$	&		&	[16]	&	$	1.06	$	&	[5]	\\
Cgy-X3	&	GBH	&	$	32.17	$	&		&	[5]	&	$	33.67	$	&		&	[17]	&	$	1.00	$	&	[5]	\\
Cgy-X3	&	GBH	&	$	32.17	$	&		&	[5]	&	$	36.37	$	&		&	[17]	&	$	1.00	$	&	[5]	\\
GRO J1655-40	&	GBH	&	$	29.94	$	&		&	[5]	&	$		$	&		&		&	$	0.85	$	&	[5]	\\
GRS 1915+105	&	GBH	&	$	31.76	$	&		&	[5]	&	$		$	&		&		&	$	1.20	$	&	[5]	\\
GX 339-4	&	GBH	&	$	29.91	$	&		&	[5]	&	$		$	&		&		&	$	1.00	$	&	[5]	\\
LS 5039	&	GBH	&	$	30.09	$	&		&	[5]	&	$		$	&		&		&	$	1.00	$	&	[5]	\\
XTE J1118+480	&	GBH	&	$	28.92	$	&		&	[5]	&	$		$	&		&		&	$	1.00	$	&	[5]	\\
XTE J1859+226	&	GBH	&	$	29.18	$	&		&	[5]	&	$		$	&		&		&	$	0.88	$	&	[5]	\\
XTE J1550-564	&	GBH	&	$	31.79	$	&		&	[6]	&	$		$	&		&		&	$	0.98	$	&	[7]	\\
 
\label{tab:radio}
\end{longtable*}

\footnotesize{\noindent Column (1): Source name;  Column (2):  source spectral class:  L - LINER; S - Seyfert; H - HII region spectral type; T: source with transition spectrum (LINER+HII); GBH: galactic black hole binary (or microquasar) \citep[for more details see][]{merloni_etal_03, nagar_etal_05}; Column (3): logarithm of the core radio luminosity (in erg/s); Column (6): logarithm of the gamma-ray luminosity (in erg/s); Column (9): logarithm of the black hole mass (in solar units); Columns (4) and (7): upper limit of the core radio and gamma luminosity (UL); Columns (5), (8) and (10): References.} 

\begin{longtable*}{lcccccccccc}
\caption{Radio and gamma-ray emission of blazars.}\\
\hline \hline \\[-2ex]
\multicolumn{1}{c}{Sources} &
\multicolumn{1}{c}{Type} &
\multicolumn{1}{c}{$\log_{10}(L_{\gamma}^{iso})$} &
\multicolumn{1}{c}{$\log_{10}(L_{\gamma})$} &
\multicolumn{1}{c}{Ref.} &
\multicolumn{1}{c}{$\log_{10}(L_R^{iso})$} &
\multicolumn{1}{c}{$\log_{10}(L_R)$} &
\multicolumn{1}{c}{Ref.} &
\multicolumn{1}{c}{$\log_{10}(M/M_{\odot})$} &
\multicolumn{1}{c}{Ref.}
\\
\multicolumn{1}{c}{(1)} &
\multicolumn{1}{c}{(2)} &
\multicolumn{1}{c}{(3)} &
\multicolumn{1}{c}{(4)} &
\multicolumn{1}{c}{(5)} &
\multicolumn{1}{c}{(6)} &
\multicolumn{1}{c}{(7)} &
\multicolumn{1}{c}{(8)} &
\multicolumn{1}{c}{(9)} &
\multicolumn{1}{c}{(10)}
\\[0.5ex] \hline
\\[-1.8ex]

\endfirsthead

\hline \hline\\[-2ex]

\multicolumn{1}{c}{Sources} &
\multicolumn{1}{c}{Type} &
\multicolumn{1}{c}{$\log_{10}(L_{\gamma}^{iso})$} &
\multicolumn{1}{c}{$\log_{10}(L_{\gamma})$} &
\multicolumn{1}{c}{Ref.} &
\multicolumn{1}{c}{$\log_{10}(L_R^{iso})$} &
\multicolumn{1}{c}{$\log_{10}(L_R)$} &
\multicolumn{1}{c}{Ref.} &
\multicolumn{1}{c}{$\log_{10}(M/M_{\odot})$} &
\multicolumn{1}{c}{Ref.}
\\
\multicolumn{1}{c}{(1)} &
\multicolumn{1}{c}{(2)} &
\multicolumn{1}{c}{(3)} &
\multicolumn{1}{c}{(4)} &
\multicolumn{1}{c}{(5)} &
\multicolumn{1}{c}{(6)} &
\multicolumn{1}{c}{(7)} &
\multicolumn{1}{c}{(8)} &
\multicolumn{1}{c}{(9)} &
\multicolumn{1}{c}{(10)}
\\[0.5ex] \hline
\\[-1.8ex]

\endhead

\multicolumn{7}{l}{{\footnotesize{Continued on next page}}}\\
\endfoot
\hline

\endlastfoot

PKS 0754+100	&	BLL	&	$	45.72	$	&	$	42.75	$	&	[18]	&	$	42.80	$	&	$	39.83	$	&	[19]	&	$	8.21	$	&	[20]	\\
PKS 0823+033	&	BLL	&	$	45.86	$	&	$	43.73	$	&	[18]	&	$	43.26	$	&	$	41.12	$	&	[19]	&	$	8.55	$	&	[20]	\\
PKS 0829+046	&	BLL	&	$	45.68	$	&	$	43.60	$	&	[18]	&	$	41.97	$	&	$	39.89	$	&	[19]	&	$	8.52	$	&	[20]	\\
OJ 287	&	BLL	&	$	46.12	$	&	$	43.89	$	&	[18]	&	$	42.82	$	&	$	40.59	$	&	[19]	&	$	8.50	$	&	[20]	\\
PKS 2155-304	&	BLL	&	$	45.98	$	&	$	43.80	$	&	[18]	&	$		$	&	$		$	&		&	$	7.60	$	&	[20]	\\
4C -02.81	&	BLL	&	$	47.18	$	&	$	44.59	$	&	[18]	&	$	44.28	$	&	$	41.69	$	&	[19]	&	$	9.96	$	&	[20]	\\
3C 454.3	&	FSRQ	&	$	48.79	$	&	$	45.91	$	&	[18]	&	$	44.85	$	&	$	41.97	$	&	[19]	&	$	6.76	$	&	[20]	\\
S3 2141+17	&	FSRQ	&	$	46.01	$	&	$	43.82	$	&	[18]	&	$		$	&	$		$	&		&	$	8.14	$	&	[20]	\\
AO 0235+164	&	BLL	&	$	47.38	$	&	$	44.72	$	&	[18]	&	$	43.98	$	&	$	41.32	$	&	[19]	&	$	7.98	$	&	[20]	\\
4C +28.07	&	FSRQ	&	$	47.78	$	&	$	45.27	$	&	[18]	&	$	44.44	$	&	$	41.92	$	&	[19]	&	$	7.98	$	&	[20]	\\
MKN 421	&	BLL	&	$	44.88	$	&	$	43.08	$	&	[18]	&	$		$	&	$		$	&		&	$	8.23	$	&	[20]	\\
PKS 1127-145	&	FSRQ	&	$	47.68	$	&	$	44.92	$	&	[18]	&	$	44.71	$	&	$	41.95	$	&	[19]	&	$	7.75	$	&	[20]	\\
4C +29.45	&	FSRQ	&	$	47.31	$	&	$	44.21	$	&	[18]	&	$	43.72	$	&	$	40.62	$	&	[19]	&	$	8.56	$	&	[20]	\\
ON 231	&	BLL	&	$	45.18	$	&	$	43.28	$	&	[18]	&	$		$	&	$		$	&		&	$	8.01	$	&	[20]	\\
4C +21.35	&	FSRQ	&	$	47.50	$	&	$	43.90	$	&	[18]	&	$	43.02	$	&	$	39.41	$	&	[19]	&	$	8.18	$	&	[20]	\\
3C 273	&	FSRQ	&	$	46.34	$	&	$	43.76	$	&	[18]	&	$	43.51	$	&	$	40.93	$	&	[19]	&	$	9.38	$	&	[20]	\\
3C 279	&	FSRQ	&	$	47.64	$	&	$	44.70	$	&	[18]	&	$	44.22	$	&	$	41.28	$	&	[19]	&	$	7.82	$	&	[20]	\\
PG 1424+240	&	BLL	&	$	45.98	$	&	$	43.80	$	&	[18]	&	$		$	&	$		$	&		&	$	6.42	$	&	[20]	\\
AP Lib	&	BLL	&	$	44.50	$	&	$	42.83	$	&	[18]	&	$		$	&	$		$	&		&	$	8.64	$	&	[20]	\\
PKS 1510-089	&	FSRQ	&	$	47.44	$	&	$	44.51	$	&	[18]	&	$	42.95	$	&	$	40.02	$	&	[19]	&	$	8.31	$	&	[20]	\\
NRAO 530	&	FSRQ	&	$	47.39	$	&	$	43.45	$	&	[18]	&	$	44.54	$	&	$	40.61	$	&	[19]	&	$	8.07	$	&	[20]	\\
OT 081	&	BLL	&	$	46.26	$	&	$	44.16	$	&	[18]	&	$	42.69	$	&	$	40.59	$	&	[19]	&	$	8.64	$	&	[20]	\\
4C +10.45	&	FSRQ	&	$	47.56	$	&	$	44.70	$	&	[18]	&	$	44.22	$	&	$	41.37	$	&	[19]	&	$	8.07	$	&	[20]	\\
3C 66A	&	BLL	&	$	47.31	$	&	$	44.67	$	&	[18]	&	$		$	&	$		$	&		&	$	8.30	$	&	[20]	\\
PKS 1604+159	&	BLL	&	$	46.06	$	&	$	43.86	$	&	[18]	&	$		$	&	$		$	&		&	$	8.25	$	&	[20]	\\
B2 1811+31	&	BLL	&	$	44.85	$	&	$	43.06	$	&	[18]	&	$		$	&	$		$	&		&	$	8.82	$	&	[20]	\\
OS 319	&	FSRQ	&	$	46.98	$	&	$	44.88	$	&	[18]	&	$	44.69	$	&	$	42.58	$	&	[19]	&	$	8.22	$	&	[20]	\\
4C +38.41	&	FSRQ	&	$	48.72	$	&	$	45.44	$	&	[18]	&	$	44.85	$	&	$	41.56	$	&	[19]	&	$	7.54	$	&	[20]	\\
MKN 501	&	BLL	&	$	44.46	$	&	$	42.80	$	&	[18]	&	$		$	&	$		$	&		&	$	8.72	$	&	[20]	\\
PKS 0454-46	&	FSRQ	&	$	47.11	$	&	$	44.54	$	&	[18]	&	$		$	&	$		$	&		&	$	8.05	$	&	[20]	\\
4C +56.27	&	BLL	&	$	46.93	$	&	$	44.61	$	&	[18]	&	$	43.40	$	&	$	41.08	$	&	[19]	&	$	9.10	$	&	[20]	\\
S5 1803+784	&	BLL	&	$	47.08	$	&	$	44.83	$	&	[18]	&	$	43.75	$	&	$	41.50	$	&	[19]	&	$	8.82	$	&	[20]	\\
BL Lac	&	BLL	&	$	45.16	$	&	$	43.40	$	&	[18]	&	$	41.50	$	&	$	39.73	$	&	[19]	&	$	8.58	$	&	[20]	\\
4C +51.37	&	FSRQ	&	$	47.83	$	&	$	45.01	$	&	[18]	&	$	44.42	$	&	$	41.61	$	&	[19]	&	$	7.97	$	&	[20]	\\
PKS 2052-47	&	FSRQ	&	$	48.33	$	&	$	45.34	$	&	[18]	&	$		$	&	$		$	&		&	$	7.88	$	&	[20]	\\
S5 0716+714	&	BLL	&	$	46.78	$	&	$	44.46	$	&	[18]	&	$	42.44	$	&	$	40.12	$	&	[19]	&	$	7.74	$	&	[20]	\\
EXO 0706.1+5913	&	BLL	&	$	44.70	$	&	$	42.96	$	&	[18]	&	$		$	&	$		$	&		&	$	8.67	$	&	[20]	\\
4C +71.07	&	FSRQ	&	$	48.36	$	&	$	45.16	$	&	[18]	&	$	45.23	$	&	$	42.03	$	&	[19]	&	$	7.22	$	&	[20]	\\
B2 0827+24	&	FSRQ	&	$	47.17	$	&	$	44.07	$	&	[18]	&	$	43.68	$	&	$	40.58	$	&	[19]	&	$	8.41	$	&	[20]	\\
MKN 180	&	BLL	&	$	43.85	$	&	$	42.40	$	&	[18]	&	$		$	&	$		$	&		&	$	8.10	$	&	[20]	\\
1ES 1028+511	&	BLL	&	$	45.76	$	&	$	43.66	$	&	[18]	&	$		$	&	$		$	&		&	$	8.70	$	&	[20]	\\
S4 0954+658	&	BLL	&	$	45.98	$	&	$	43.80	$	&	[18]	&	$		$	&	$		$	&		&	$	8.37	$	&	[20]	\\
1ES 0806+524	&	BLL	&	$	45.14	$	&	$	43.25	$	&	[18]	&	$		$	&	$		$	&		&	$	8.65	$	&	[20]	\\
4C +55.17	&	FSRQ	&	$	47.70	$	&	$	44.93	$	&	[18]	&	$		$	&	$		$	&		&	$	8.42	$	&	[20]	\\
PG 1246+586	&	BLL	&	$	47.20	$	&	$	44.60	$	&	[18]	&	$		$	&	$		$	&		&	$	9.15	$	&	[20]	\\
S4 0814+425	&	BLL	&	$	46.17	$	&	$	45.01	$	&	[18]	&	$	42.41	$	&	$	41.24	$	&	[19]	&	$	8.01	$	&	[20]	\\
S4 0917+44	&	FSRQ	&	$	48.72	$	&	$	45.60	$	&	[18]	&	$		$	&	$		$	&		&	$	7.88	$	&	[20]	\\
PG 1437+398	&	BLL	&	$	45.31	$	&	$	43.36	$	&	[18]	&	$		$	&	$		$	&		&	$	8.95	$	&	[20]	\\
PKS 0336-019	&	FSRQ	&	$	46.96	$	&	$	43.93	$	&	[18]	&	$	44.16	$	&	$	41.13	$	&	[19]	&	$	7.21	$	&	[20]	\\
PKS 0420-01	&	FSRQ	&	$	47.62	$	&	$	45.22	$	&	[18]	&	$	44.24	$	&	$	41.84	$	&	[19]	&	$	8.04	$	&	[20]	\\
PKS 0440-00	&	FSRQ	&	$	47.57	$	&	$	44.84	$	&	[18]	&	$		$	&	$		$	&		&	$	7.23	$	&	[20]	\\
4C -02.19	&	FSRQ	&	$	48.02	$	&	$	45.28	$	&	[18]	&	$	44.98	$	&	$	42.24	$	&	[19]	&	$	8.50	$	&	[20]	\\
PKS 0528+134	&	FSRQ	&	$	48.32	$	&	$	45.36	$	&	[18]	&	$	45.00	$	&	$	42.04	$	&	[19]	&	$	7.66	$	&	[20]	\\
PKS 0537-441	&	BLL	&	$	48.25	$	&	$	45.29	$	&	[18]	&	$		$	&	$		$	&		&	$	8.02	$	&	[20]	\\
PKS 0735+17	&	BLL	&	$	46.57	$	&	$	44.19	$	&	[18]	&	$	43.23	$	&	$	40.85	$	&	[19]	&	$	8.08	$	&	[20]	\\
PKS 2201+04	&	BLL	&	$	43.11	$	&	$	41.91	$	&	[18]	&	$		$	&	$		$	&		&	$	7.76	$	&	[20]	\\
1ES 1741+196	&	BLL	&	$	44.18	$	&	$	42.62	$	&	[18]	&	$		$	&	$		$	&		&	$	8.93	$	&	[20]	\\

\label{tab:blazar}
\end{longtable*}

\footnotesize{ \noindent  Column (1): source name;  Column (2): BLL: BL Lac objects; FSRQ: Flat Spectrum Radio Quasar; Column (3): logarithm of the isotropic gamma-ray luminosity (in erg/s); Column (4): logarithm of the gamma-ray luminosity (in erg/s) corrected by Doppler boosting; Column (6): logarithm of the isotropic radio luminosity (in erg/s); Column (7): logarithm of the radio luminosity (in erg/s) corrected by Doppler boosting; Column (9): logarithm of the black hole mass (in solar units); Columns (5), (8) and (10): References. }

\begin{longtable*}{lcccccccc}
\caption{Gamma-ray emission of GRBs.}\\
\hline \hline \\[-2ex]
\multicolumn{1}{c}{Sources} &
\multicolumn{1}{c}{$\log_{10}(L_{\gamma}^{iso}[erg/s])$} &
\multicolumn{1}{c}{$\log_{10}(L_{\gamma}[erg/s])$} &
\multicolumn{1}{c}{Ref.} &
\multicolumn{1}{c}{Sources} &
\multicolumn{1}{c}{$\log_{10}(L_{\gamma}^{iso}[erg/s])$} &
\multicolumn{1}{c}{$\log_{10}(L_{\gamma}[erg/s])$} &
\multicolumn{1}{c}{Ref.}
\\
\multicolumn{1}{c}{(1)} &
\multicolumn{1}{c}{(2)} &
\multicolumn{1}{c}{(3)} &
\multicolumn{1}{c}{(4)} &
\multicolumn{1}{c}{(5)} &
\multicolumn{1}{c}{(6)} &
\multicolumn{1}{c}{(7)} &
\multicolumn{1}{c}{(8)}
\\[0.5ex] \hline
\\[-1.8ex]

\endfirsthead

\hline \hline\\[-2ex]

\multicolumn{1}{c}{Sources} &
\multicolumn{1}{c}{$\log_{10}(L_{\gamma}^{iso}[erg/s])$} &
\multicolumn{1}{c}{$\log_{10}(L_{\gamma}[erg/s])$} &
\multicolumn{1}{c}{Ref.} &
\multicolumn{1}{c}{Sources} &
\multicolumn{1}{c}{$\log_{10}(L_{\gamma}^{iso}[erg/s])$} &
\multicolumn{1}{c}{$\log_{10}(L_{\gamma}[erg/s])$} &
\multicolumn{1}{c}{Ref.}
\\
\multicolumn{1}{c}{(1)} &
\multicolumn{1}{c}{(2)} &
\multicolumn{1}{c}{(3)} &
\multicolumn{1}{c}{(4)} &
\multicolumn{1}{c}{(5)} &
\multicolumn{1}{c}{(6)} &
\multicolumn{1}{c}{(7)} &
\multicolumn{1}{c}{(8)}
\\[0.5ex] \hline
\\[-1.8ex]

\endhead

\multicolumn{7}{l}{{\footnotesize{Continued on next page}}}\\
\endfoot
\hline

\endlastfoot

90323	&	$	53.06	$	&	$	50.14	$	&	[18]	&	21004	&	$	51.55	$	&	$	49.93	$	&	[18]	\\
90328	&	$	51.47	$	&	$	48.9	$	&	[18]	&	31203	&	$	49.43	$	&	$	47.52	$	&	[18]	\\
090902B	&	$	53.63	$	&	$	51	$	&	[18]	&	30329	&	$	50.82	$	&	$	48.42	$	&	[18]	\\
090926A	&	$	53.46	$	&	$	51.55	$	&	[18]	&	50709	&	$	51.22	$	&	$	49.87	$	&	[18]	\\
81222	&	$	53.28	$	&	$	50.35	$	&	[18]	&	050820A	&	$	53.13	$	&	$	50.95	$	&	[18]	\\
90424	&	$	51.52	$	&	$	49.69	$	&	[18]	&	50904	&	$	52.63	$	&	$	50.62	$	&	[18]	\\
90618	&	$	52.19	$	&	$	50.02	$	&	[18]	&	60218	&	$	46.09	$	&	$	46.01	$	&	[18]	\\
91020	&	$	51.65	$	&	$	49.51	$	&	[18]	&	60418	&	$	51.28	$	&	$	50.16	$	&	[18]	\\
91127	&	$	50.47	$	&	$	48.13	$	&	[18]	&	70125	&	$	52.61	$	&	$	51.03	$	&	[18]	\\
091208B	&	$	50.8	$	&	$	48.7	$	&	[18]	&	080319B	&	$	51.26	$	&	$	49.13	$	&	[18]	\\
970228	&	$	50.63	$	&	$	48.02	$	&	[18]	&	50505	&	$	51.86	$	&	$	48.66	$	&	[18]	\\
970508	&	$	50.86	$	&	$	49.71	$	&	[18]	&	50814	&	$	52.36	$	&	$	49.32	$	&	[18]	\\
970828	&	$	51.43	$	&	$	49.32	$	&	[18]	&	051109A	&	$	50.79	$	&	$	48.02	$	&	[18]	\\
971214	&	$	52.49	$	&	$	50.16	$	&	[18]	&	051221A	&	$	50.7	$	&	$	49.02	$	&	[18]	\\
980613	&	$	50.43	$	&	$	48.81	$	&	[18]	&	60124	&	$	53.6	$	&	$	50.55	$	&	[18]	\\
980425	&	$	46.84	$	&	$	45.1	$	&	[18]	&	60614	&	$	49.73	$	&	$	48.05	$	&	[18]	\\
980703	&	$	51.19	$	&	$	49.47	$	&	[18]	&	60707	&	$	51.18	$	&	$	49.16	$	&	[18]	\\
990123	&	$	52.79	$	&	$	50.36	$	&	[18]	&	60814	&	$	52.96	$	&	$	50.24	$	&	[18]	\\
990510	&	$	51.91	$	&	$	49.14	$	&	[18]	&	61021	&	$	49.75	$	&	$	47.8	$	&	[18]	\\
990705	&	$	52.17	$	&	$	49.8	$	&	[18]	&	061222A	&	$	52.76	$	&	$	49.8	$	&	[18]	\\
991216	&	$	52.86	$	&	$	50.36	$	&	[18]	&	70306	&	$	52.7	$	&	$	50.16	$	&	[18]	\\
21004	&	$	52.54	$	&	$	50.39	$	&	[18]	&	70318	&	$	52.69	$	&	$	50.76	$	&	[18]	\\
926	&	$	53.82	$	&	$	51.58	$	&	[18]	&	70508	&	$	51.76	$	&	$	49.03	$	&	[18]	\\
10222	&	$	52.46	$	&	$	49.65	$	&	[18]	&	80310	&	$	51.81	$	&	$	49.1	$	&	[18]	\\
11211	&	$	51.62	$	&	$	49.41	$	&	[18]	&	080413B	&	$	52.61	$	&	$	50.35	$	&	[18]	\\
20405	&	$	51.49	$	&	$	49.45	$	&	[18]	&	90313	&	$	51.23	$	&	$	48.38	$	&	[18]	\\
20813	&	$	52.29	$	&	$	49.47	$	&	[18]	&	91018	&	$	50.04	$	&	$	47.57	$	&	[18]	\\

\label{tab:grb}
\end{longtable*}

\footnotesize{ \noindent Columns (1) and (5): source name; Columns (2) and (6) logarithm of the isotropic gamma-ray luminosity (in erg/s);  Column (3) and (7): logarithm of the gamma-ray luminosity (in erg/s) corrected by Doppler boosting; Columns (4) and (8): References.}

\footnotesize{ \noindent Notes. References (for all Tables): [0] \citet{nagar_etal_05}, [1] \citet{richstone_etal_98}, [2] \citet{gebhardt_etal_00}, [3] \citet{merritt_ferrarese_01}, [4] \citet{tremaine_etal_02}, [5] \citet{merloni_etal_03}, [6] \citet{hannikainen_etal_01}, [7] \citet{remillard_mcclintock_06}, [8] \citet{ackermann_etal_12}, [9] \citet{middleton_etal_08}, [10] \citet{abdo_etal_09b}, [11] \citet{israel_98}, [12] \citet{abdo_etal_10}, [13] \citet{aleksic_etal_14b}, [14] \citet{aleksic_etal_14a}, [15] \citet{kadler_etal_12}, [16] \citet{malyshev_etal_13}, [17] \citet{piano_etal_12}, [18] \citet{nemmen_etal_12}, [19] \citet{kharb_etal_10} and [20] \citet{vovk_neronov_13}.}

\end{appendix}

\end{document}